\documentclass[fleqn,10pt]{wlscirep}
\usepackage[utf8]{inputenc}
\usepackage[T1]{fontenc}
\usepackage{amsmath}
\usepackage{nccmath}
\usepackage{bibunits}
\usepackage{url}

\graphicspath{{figures/}}

\title{
One City, Two Tales: Using Mobility Networks to Understand Neighborhood Resilience and Fragility during the COVID-19 Pandemic\\
}

\author[1,+]{Hasan Alp Boz}
\author[2,*,+]{Mohsen Bahrami}
\author[1]{Selim Balcisoy}
\author[3]{Burcin Bozkaya}
\author[4]{Nina Mazar}
\author[4]{Aaron Nichols}
\author[2]{Alex Pentland}
\affil[1]{Sabanci University, Faculty of Engineering and Natural Sciences, Istanbul, 34956, Turkey}
\affil[2]{Massachusetts Institute of Technology, Institute for Data, Systems, and Society, Cambridge, MA 02139, USA}
\affil[3]{New College of Florida, The Graduate Program in Applied Data Science, 5800 Bay Shore Rd, Sarasota, FL 34243, USA}
\affil[4]{Boston University, Questrom School of Business, Boston, MA 02215, USA}

\affil[*]{bahrami@mit.edu}

\affil[+]{These authors contributed equally to this work. Authors are listed alphabetically.}

\begin{abstract}
What predicts a neighborhood’s resilience and adaptability to essential public health policies and shelter-in-place regulations that prevent the harmful spread of COVID-19? To answer this question, in this paper we present a novel application of human mobility patterns and human behavior in a network setting. We analyze mobility data in New York City over two years, from January 2019 to December 2020, and create weekly mobility networks between Census Block Groups by aggregating Point of Interest level visit patterns. Our results suggest that both the socioeconomic and geographic attributes of neighborhoods significantly predict neighborhood adaptability to the shelter-in-place policies active at that time. That is, our findings and simulation results reveal that in addition to factors such as race, education, and income, geographical attributes such as access to amenities in a neighborhood that satisfy community needs were equally important factors for predicting neighborhood adaptability and the spread of COVID-19. The results of our study provide insights that can enhance urban planning strategies that contribute to pandemic alleviation efforts, which in turn may help urban areas become more resilient to exogenous shocks such as the COVID-19 pandemic.

\end{abstract}
\begin{document}
\flushbottom
\maketitle
%
%
\thispagestyle{empty}


\section*{Introduction}

Mobility in urban and metropolitan settings is the result of the dynamic interaction, over space and time, of a large number of agents with diverse goals and characteristics. Understanding mobility patterns is crucial for predicting future social and economic well-being as well as growth \cite{chong2020economic, singh2015money}, as it has been shown to reflect social interactions \cite{bettencourt2013origins, buera2020global}, productivity \cite{sveikauskas1975productivity, bettencourt2007growth}, and economic prosperity and resilience \cite{alvarez2013idea, steele2017mapping, pan2013urban, schlapfer2021universal}. 

The dynamic interactions in mobility data are complex and are typically represented by network structures. However, despite the ability for network analyses to yield unique and critically important insights \cite{lazer2009life}, researchers have yet to use network analyses to explore how the COVID-19 pandemic and the shelter-in-place and social distancing policies that it triggered -- two of the most effective non-pharmaceutical interventions (NPI) aimed at reducing the spread of the Corona virus \cite{flaxman2020estimating, kraemer2020effect} (for a differing conclusion, see a recently published article by Berry et al.\cite{berry2021evaluating}) -- influenced human mobility and interaction patterns. Such interventions directly impact the human mobility network by reducing levels of mobility and interactions among individuals and points of interest (POIs) such as restaurants or supermarkets \cite{aleta2020modelling, schlosser2020covid}. The decline in movement activities during the current COVID-19 pandemic has negatively impacted economies all around the world, and these economic repercussions are predicted to continue for years \cite{chetty2020economic}. 

Although network analysis approaches are well suited to investigate the multifaceted impact that policy changes have on mobility outcomes, early studies have examined the short-term impacts of the COVID-19 pandemic on various unidimensional mobility metrics such as travel distance, visit patterns, and dwelling times in different countries and cities \cite{gao2020mapping, galeazzi2020human}, or used  mobility networks along with epidemiological models to understand the effects of changes in mobility patterns on the spread of the Corona virus \cite{chang2021mobility, aleta2020modelling}. Indeed, there is a lack of research investigating how the COVID-19 pandemic and corresponding interventions impacted the structure of mobility networks (e.g. the centrality metrics at a node and network level). For example, investigating the \emph{betweenness} centrality metric may reveal the role certain neighborhoods played during the pandemic, such as acting as a bridge among the nodes in a mobility network, thereby producing a spreader effect.

Given the close link between changes in mobility patterns and economic outcomes, it is critical that research investigates how various environmental and demographic factors have influenced the adaptability of mobility networks during the COVID-19 pandemic. Using network science methodologies, the current research aims to help scientists and policy-makers understand the factors that contribute to economic resilience and adaptability in the wake of economic shocks. In particular, we extend previous research about the impact of the COVID-19 pandemic on human behavior by examining changes in mobility patterns using a dynamic network analysis on one of the most important economic hubs in the world: New York City (NYC). We create weekly mobility networks, from January 2019 to December 2020, in which nodes represent neighborhoods (i.e. Census Block Groups or CBGs), and the edges between them correspond to the visitors from the source neighborhood to POIs such as restaurants or supermarkets in the target neighborhood. For each neighborhood in the weekly networks, we compute node and ego-network based features \cite{netsimile} so the resulting feature vectors not only capture the dynamics of local mobility but also the relationship with the neighboring CBGs, allowing us to incorporate the complexity of mobility into our analyses. We investigate the dissimilarity of the resulting feature vectors for each neighborhood, between the same weeks of 2019 and 2020, and break the results down by different socioeconomic groups. Combining the mobility network metrics with data from various sources, including census data and COVID-19 test results, we are able to reveal how differences in NYC neighborhood characteristics predict dynamic structural changes of mobility networks and behavior.

The results of our study indicate that the centrality metrics and geographic attributes significantly predict neighborhood adaptability to shelter-in-place orders. In addition to confirming the results of previous research \cite{chang2021mobility, chetty2020economic, hunter2021effect, heroy2021covid}, our findings reveal that not only are race, education, and income important factors in predicting neighborhood adaptability to shelter-in-place orders, but so are geographical attributes such as access to diverse amenities that satisfy community needs. This indicates that in the same city, communities with similar socioeconomic and demographic features may have different mobility responses based on their neighborhoods' urban structure. 

Using the information extracted from the mobility network structure, we study the case of COVID-19 hotspots to investigate which neighborhoods act as the COVID-19 bridges among the hotspots and other neighborhoods, and uncover the associated factors. We then utilize the well-known Huff gravity model to perform a hypothetical scenario analysis to show how the higher levels of access to essential businesses (e.g. grocery stores) could reduce the interaction among the COVID-19 hotspots and other neighborhoods that could potentially lead to a reduction in the infection rates and save more lives. These novel results provide significant insights and recommend policies that can enhance urban planning strategies that contribute to pandemic alleviation efforts, which in turn may help urban areas become more resilient to exogenous shocks such as the COVID-19 pandemic that restrict movements and interactions.



\section*{Results} 

To demonstrate the impact of the COVID-19 pandemic on different socioeconomic groups, we first analyze the change in network topologies in a weekly resolution at the neighborhood level. To this end, we extract the node-level feature vectors summarizing the statistical properties of their respective ego-networks \cite{netsimile}.
The resulting node feature vectors are used to compute the dissimilarities between paired weekly networks from 2019 and 2020. Then, we illustrate the course of centrality metrics with respect to different demographic groups and highlight their variability. Next, we analyze the possible COVID-19 bridges, neighborhoods that frequently interact with COVID-19 hotspots (CBGs with higher numbers of infected residents), by focusing on the in- and out-going edges between CBGs over two-week periods, which is considered as the virus incubation time \cite{chang2021mobility}. Finally, utilizing the Huff Gravity Model, we analyze the mobility in Staten Island using hypothetical Grocery Store densities in order to observe the change in visits to hotspot CBGs that frequently appear in the top new COVID-19 cases quartile. 

\subsection*{Demographic Disparities: Temporal Changes in Mobility Networks}

\subsubsection*{CBG-level dissimilarity analysis}

\begin{figure}[ht!]
\centering
\includegraphics[width=\linewidth]{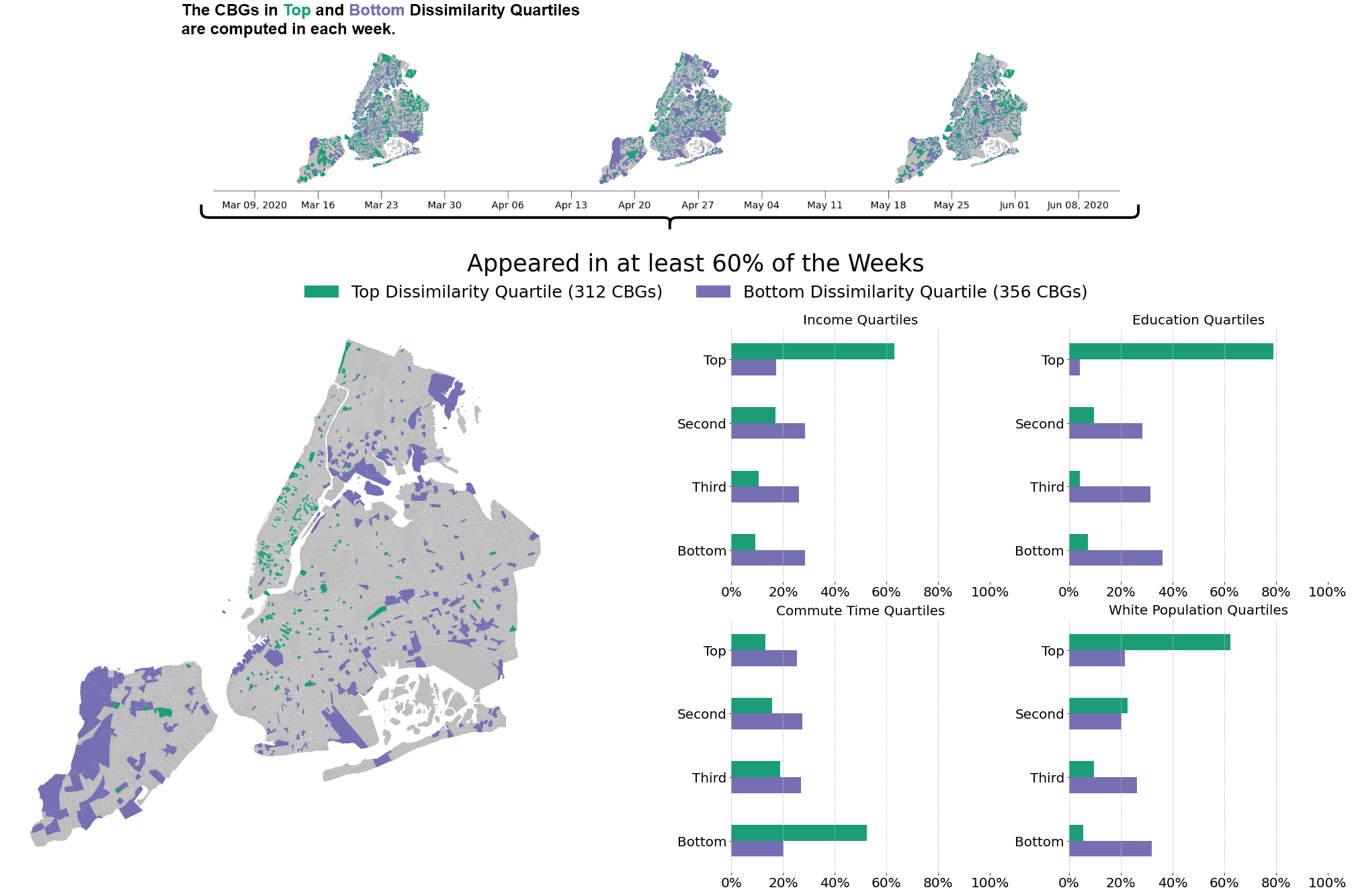}
\caption{The socioeconomic distribution of the census block groups (CBGs) that changed their mobility patterns the most in comparison to the previous year in at least 60\% of the time steps, versus the least. CBGs that ended up changing their mobility patterns the most are primarily located in the financial center of NYC. Note that there are only significant socioeconomic characteristics for the top quartiles but not for the bottom.}
\label{fig:cbg-disparity}
\end{figure}

We compute the node-level dissimilarity scores between paired weekly networks of 2019 (pre-pandemic) and 2020 (pandemic) using the extracted ego-network feature vectors (the components of these feature vectors and dissimilarity score formulation are explained in the Methods section). The CBGs are then ranked with respect to their dissimilarity scores at each time step (week). To demonstrate the differences between the CBGs with distinct behaviors, we focus on the CBGs in the top and bottom dissimilarity quartiles, and create two cohorts of CBGs that frequently appear in those quartiles during the first wave of the pandemic between March and June 2020. Figure \ref{fig:cbg-disparity} shows the spatial and socioeconomic distribution of the resulting cohorts of CBGs that appear in at least 60\% of the time steps in the top and bottom quartiles. This threshold (60\%) is the highest frequency that yields a similar number of CBGs in each group, enabling a better comparison between them. CBGs that ended up changing their mobility pattern the most (i.e. the top dissimilarity quartile) are primarily located in the Manhattan borough, the financial center of NYC.
From all CBGs in this cohort, 63\% of them rank in the top quartile for income, 79\% in the top education quartile, 62\% in the top white population percentage quartile, and 52\% in the bottom quartile for commute time, meaning they either do not travel relatively long distances to get to their workplace or are located in areas that have greater access to fast and frequent transportation. There is no evident socioeconomic profile for the bottom dissimilarity quartile (the CBGs that did not change their mobility patterns much), although the distributions in quartiles delineate the residents to some extent. However, there exists a decreasing trend from bottom to top quartiles in income, education, and white population percentage. 

\subsubsection*{Node Degree and Centrality Metrics}
Centrality metrics help us examine a node's role in the network, such as influence and information diffusion \cite{deville2016scaling, miritello2011dynamical}. In the proposed mobility network, since each node is a CBG, the centrality metrics highlight the CBGs that are noteworthy in terms of the flow of masses. 

In this context, the temporal changes in centrality metrics reveal the interaction patterns between different socioeconomic communities, which consequently indicate complex mobility behaviors from a network perspective. To this end, we focus on basic node degree centrality metrics and analyze their course of change in distinct demographic groups (i.e. CBGs in the top and bottom quartiles). We use the node \emph{betweenness} to illustrate a CBG's importance based on its connections and position in the network. Furthermore, we use \emph{degree} centrality metrics to reveal the weekly incoming and outgoing visitors among CBGs. Lastly, we use a custom metric named \emph{self-visit ratio} to represent the fraction of visits to the POIs inside the home CBG.

\begin{figure}[ht!]
\centering
\includegraphics[width=0.8\linewidth]{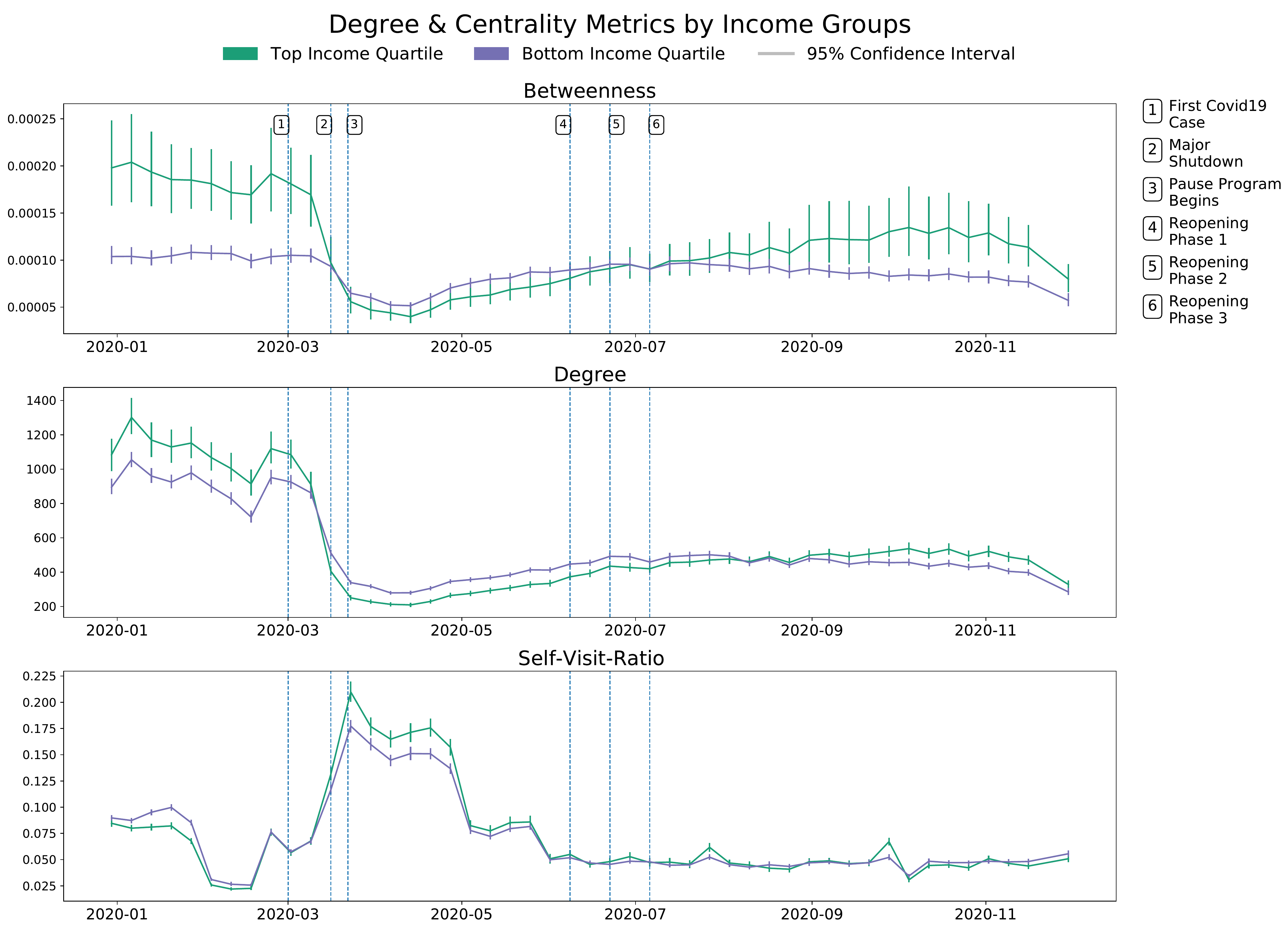}
\caption{The temporal change in (A) betweenness, (B) total-degree, and (C) self-visit ratio metrics in the top and bottom income quartiles. The vertical line segments show a 95\% confidence interval.}
\label{fig:income-cent-change}
\end{figure}

\textbf{\emph{Betweenness}:} This centrality score measures how frequently a CBG appears along the shortest paths in a network, and is the only node centrality metric that demonstrates a significant difference between the selected demographic groups. As displayed in Figure \ref{fig:income-cent-change}-A, CBGs in the top income quartile held a higher betweenness value until the beginning of March 2020 (start of the pandemic), meaning that they played a critical role in terms of bridging the flow of masses. However, an abrupt decrease of betweenness in the top income CBGs took place after the start of the pandemic, while less affluent CBGs gained higher betweenness scores. That is, less affluent CBGs increasingly acted as connectors among the nodes in the mobility network but only until September 2020, when the economic activity revived. The same relationship can also be observed when focusing on education levels. CBGs with lower education levels had a higher betweenness score in the same time interval (displayed in the Supplementary Information (SI) Figure 4). 

\textbf{\emph{Degree}:} Node degree analysis indicates that income and education play a significant role in distribution of degree centrality values as well. Affluent CBGs were more successful at lowering their mobility compared to less affluent neighborhoods as shown in Figure \ref{fig:income-cent-change}-B. 

\textbf{\emph{Self-Visit Ratio}:} Another metric we defined to investigate the change in visit patterns is called self-visit ratio. Self-visit ratio is the fraction of the visits made by the residents of a CBG to their home CBG over all visits paid by them. 
Figure \ref{fig:income-cent-change}-C displays the course of the self-visit ratio with respect to the top and bottom income quartiles. From March to June 2020, during the first wave and the most striking decline in mobility, CBGs in the top income quartile had a higher rate of visits to the POIs inside their home CBGs, while on the contrary, the residents of lower income CBGs displayed a lower self-visit ratio. However, starting in June 2020, as the re-opening of economic activity commenced, the disparity of self-visit ratio between income groups is narrowed.

\subsection*{Analysis of COVID-19 Hotspots, Bridge CBGs \& the Case of Staten Island}
As explained in the Methods section, we define the COVID-19 hotspots as those CBGs that frequently appear among the CBGs with highest weekly new cases. Additionally, we refer to the CBGs that have a high level of interactions with hotspots in the beginning of the virus incubation time as COVID-19 bridge CBGs. Examining the COVID-19 bridge CBGs may reveal invaluable insights for policy makers and urban planners attempting to prevent the spread of new infections and build cities that are resilient to future pandemics.  

To this end, we first obtain the CBGs in the top weekly new cases quartile for each time step $t$. Then, we create a list of CBGs that had edges to the previously obtained CBGs with the highest cases in time step $t-2$ considering a period of two weeks as the incubation time for new cases to surface \cite{chang2021mobility}.
Subsequently, we apply a frequency analysis on the possible bridges to check how often they were connected to CBGs with the highest weekly cases, and finally, consider the CBGs in the $75^{th}$ frequency percentiles as bridges, for further analyses.

\begin{figure}[ht!]
\centering
\includegraphics[width=\linewidth]{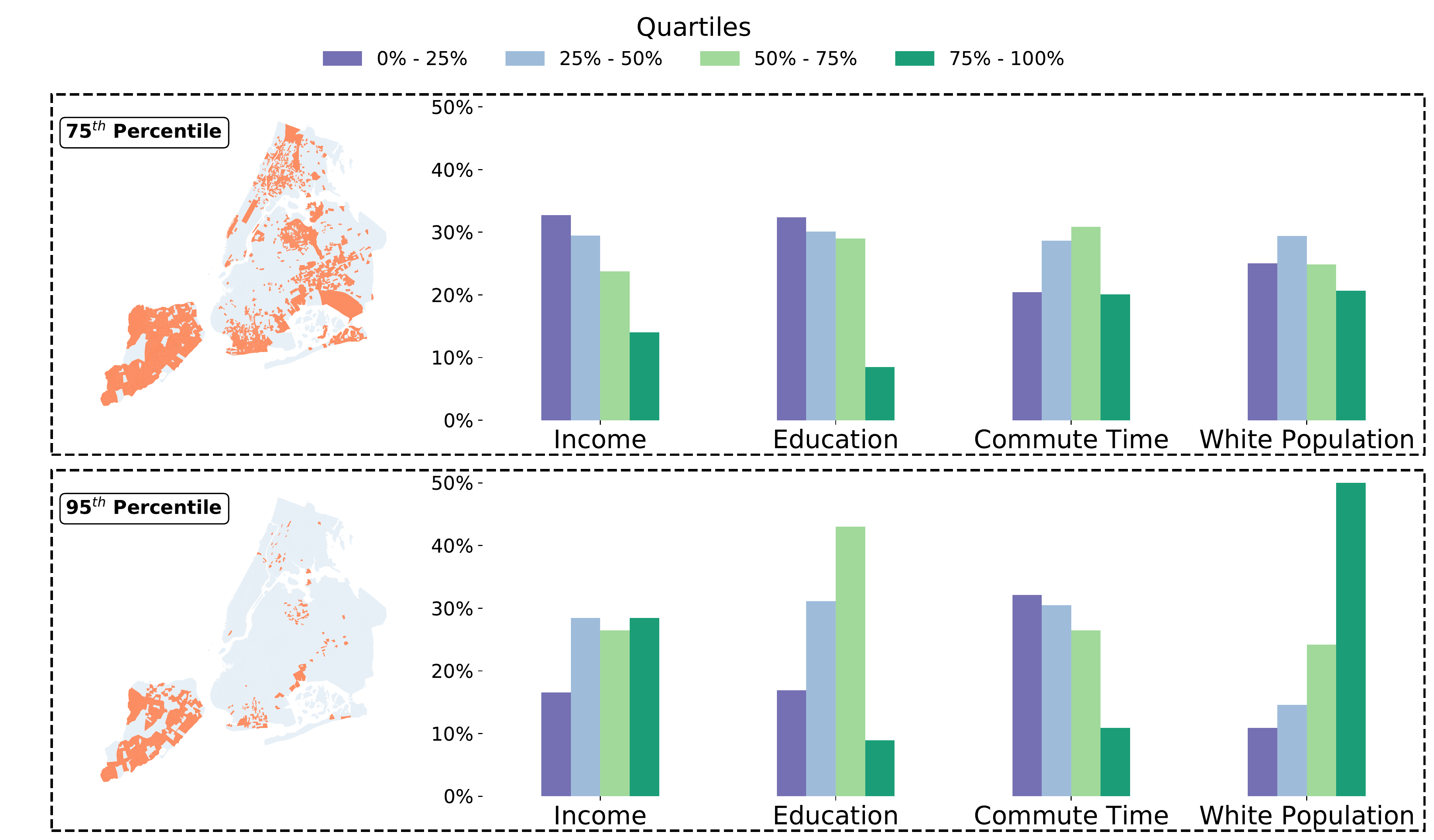}
\caption{Spatial and demographic distributions of the CBGs in the 75th (top) and 95th (bottom) frequency percentiles as COVID-19 bridges. Staten Island clearly stands out.}
\label{fig:hotspots}
\end{figure}

\begin{figure}[ht!]
\centering
\includegraphics[width=0.5\linewidth]{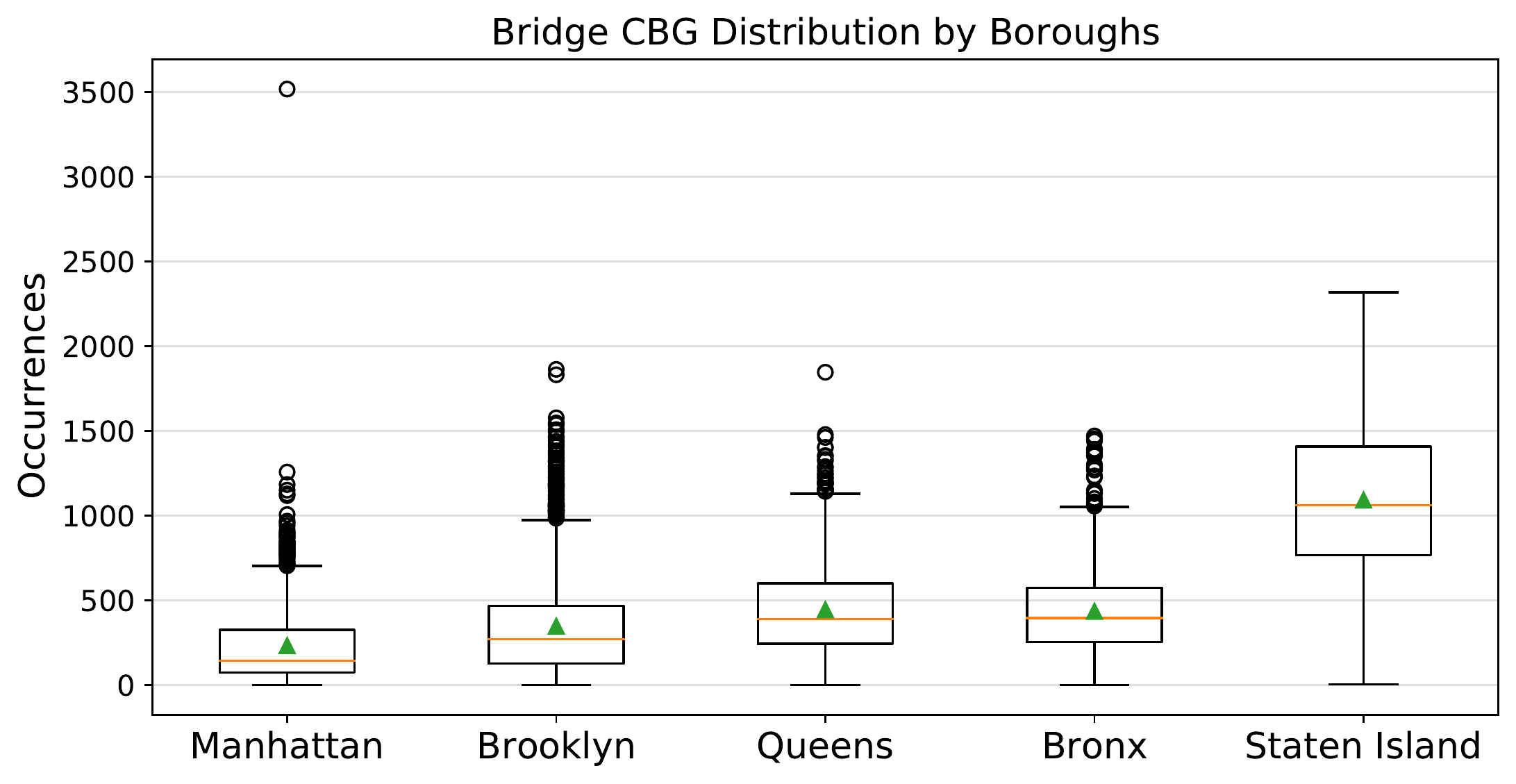}
\caption{Box plot of the COVID-19 bridge CBGs by boroughs}
\label{fig:boxplot}
\end{figure}

As demonstrated in Figure \ref{fig:hotspots}, the majority of the resulting CBGs in the $75^{th}$ frequency percentile are comprised of those in the lower quartiles for income and education and higher quartiles for commuting time. Yet the spatial distribution of the potential bridges in the $75^{th}$ frequency percentile in Figure \ref{fig:hotspots} unveils the special case of Staten Island, where 85\% of the CBGs located in Staten Island appeared in the bridges. Moreover, as the threshold value is increased to the $95^{th}$ frequency percentile, the demographic features begin to display Staten Island's presence. Figure \ref{fig:boxplot} shows the box plot of the COVID-19 bridge CBGs at a borough level. Additionally, the results of our OLS regression analysis considering occurrence in the bridge CBG set as a dependent variable and borough code as an independent variable, showed the boroughs are significant in predicting occurrence of CBGs in the bridge set (the regression analysis results are provided in SI). This is counter-intuitive and does not align with the previous observations, because 48\% of the CBGs in Staten Island are from high income quartiles and their residents are mainly white. That is, the CBGs in Staten Island display a distinctive behavior compared to CBGs in other boroughs that are similarly in high income and high white population quartiles. This observation is particularly noteworthy since Staten Island is geographically relatively isolated with a low level of connectivity to other boroughs (connected through bridges and ferry; and not connected with the NYC subway system). Thus, if anything, one may conclude the opposite, that is, Staten Island would have been more shielded from the pandemic. Therefore, we followed up with additional analysis at the borough level to try to shed light on this observation. 

\subsubsection*{Borough Level Analyses Results}

\begin{figure}[ht!]
\centering
\includegraphics[width=\linewidth]{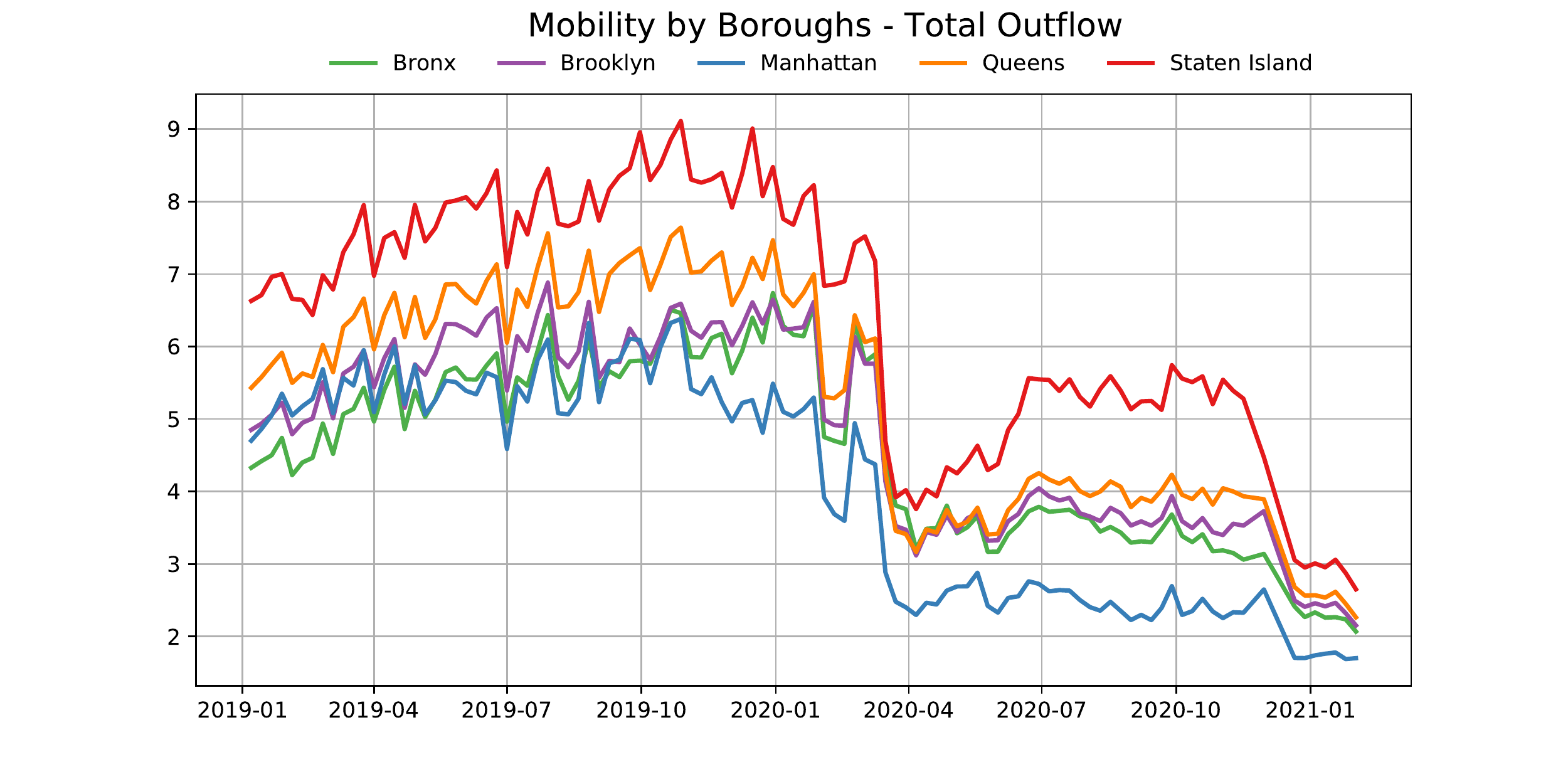}
\caption{Average weekly number of visits per user at a borough level}
\label{fig:boroughmobility}
\end{figure}

As depicted in Figure \ref{fig:boroughmobility} from January 2019 to January 2021, residents in Staten Island always had the highest mobility (average visit count per smartphone user) among all NYC boroughs, regardless of the COVID-19 pandemic and the rise of different containment policies (e.g. shelter-in-place or physical/social distancing).

\begin{figure}[ht!]
\centering
\includegraphics[width=\linewidth]{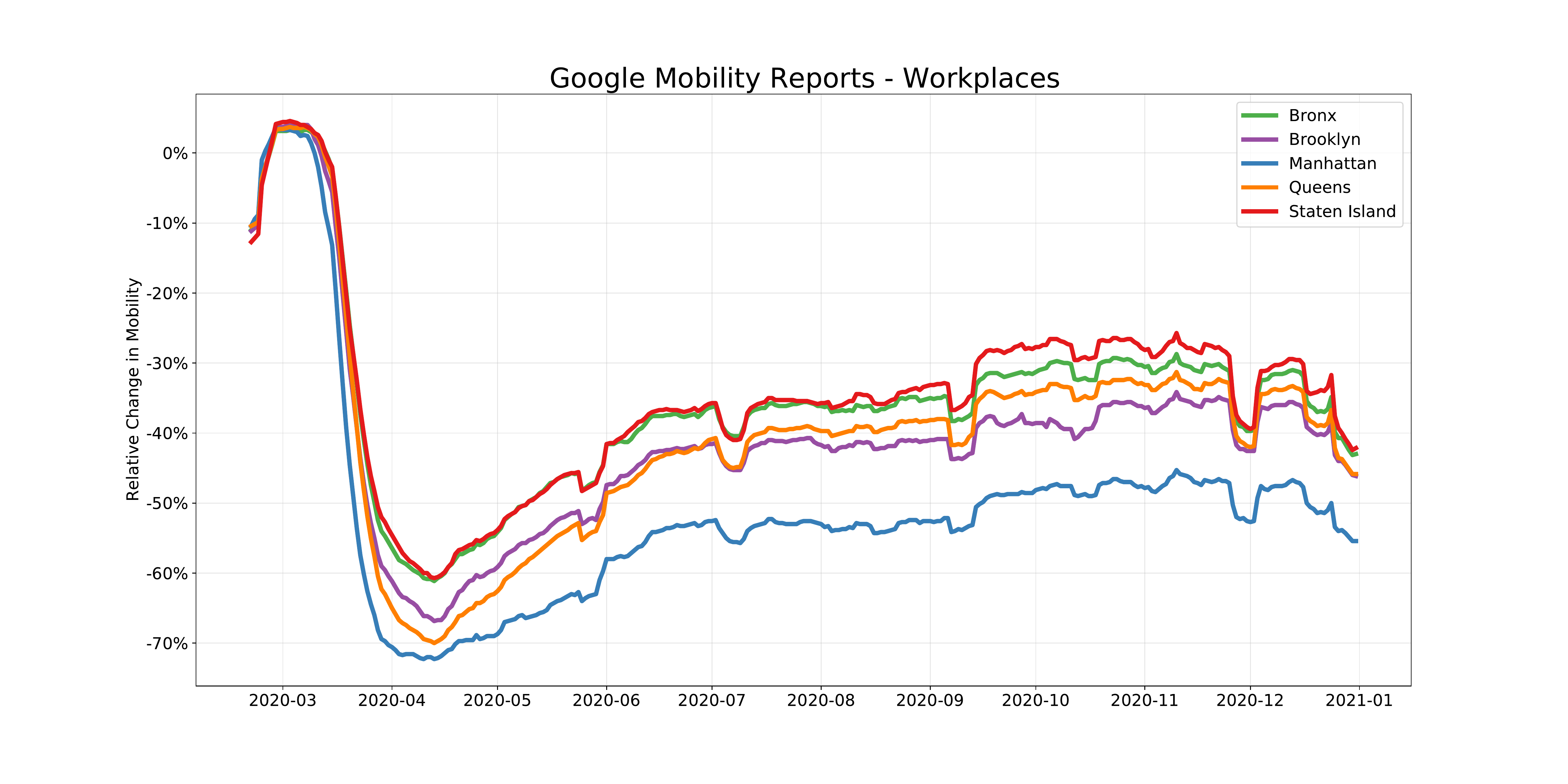}
\caption{Relative change in mobility trends for workplace by borough.}
\label{fig:googlemobility}
\end{figure}

Since the Safegraph mobility data has limited coverage on workplaces and offices, we use Google's COVID-19 Community Mobility Reports \cite{googlemobility} to investigate the mobility trends for places of work. As shown in Figure \ref{fig:googlemobility} Staten Island has the minimum relative change in mobility trends for workplace among all NYC boroughs, indicating that the residents of Staten Island reduced their mobility and visits to POIs noticeably less than the residents of other boroughs. 

Additionally, the POIs analysis results show that Staten Island has the lowest number and diversity of POIs among all boroughs of NYC, where the majority of visits to POIs inside the city originating from Staten Island are made to Brooklyn and Manhattan, which are the neighboring boroughs of Staten Island. This observation is in line with a report by the NYC government's planning department \cite{nycinsouts} documenting that 24\% of workers residing in Staten Island have their workplaces located in Manhattan.

In summary, in Staten Island due to the shortage of POIs and its relatively isolated location (e.g., just one automobile bridge to Brooklyn, three automobile bridges to New Jersey, one free ferry to Manhattan, no subway), which promotes using personal cars\cite{nycmobilityreport , nycinsouts}, the residents had to visit POIs in other boroughs to accommodate their needs and thus traveled longer distances to workplaces and the majority of POI categories. We conclude that this is likely why we observe the CBGs in Staten Island displayed a distinct response behavior compared to their demographic counterparts (i.e. relatively high income and percent white population CBGs), and that this is what led to higher infection rates in its neighborhoods. 


\subsection*{Hypothetical Scenario Analysis}

In order to understand the mobility patterns of CBGs in Staten Island under hypothetical POI distributions, we employ the Huff Gravity Model \cite{huff1964defining} focusing on the distance between CBGs and POIs, and POI areas in square feet (more details are provided in SI). In addition, instead of including all POI categories, we only incorporated grocery stores into the model to narrow down the scope to the stores that provide the most essential needs of human daily life. Table \ref{grocery_distribution} provides information about the number of grocery stores per 1K residents and the median distance travelled by residents to visit grocery stores in kilometers at a borough level.

\begin{table}[hbt!]
\centering
\begin{tabular}{l|c|c}
\hline
\textbf{Borough Name} & \textbf{Grocery Stores per 1K Residents} & \textbf{Median Distance Travelled (in km)}\\
\hline
Manhattan & 0.582 & 0.90\\
Brooklyn & 0.470 & 1.35\\
Bronx & 0.435 & 1.26\\
Queens & 0.414 & 1.71\\
Staten Island & 0.332 & 2.66\\
\hline
\end{tabular}
\caption{Number of grocery stores per 1K residents and the median distance travelled from home to grocery stores (from March 22nd and June 8th 2020) by residents of NYC boroughs.}
\label{grocery_distribution}
\end{table}

As shown in the table, Staten Island has the lowest number of grocery stores per resident and highest distance travelled by residents to visit a grocery store among all NYC boroughs. Our aim is to simulate the mobility patterns of Staten Island residents with a higher POI density equal to that of other boroughs with similar demographic traits. We choose the analysis time frame to be during the first wave of the pandemic between March 22nd (pause program begins) and June 8th (phase I of the reopening) of 2020. Since NYC enforced a strict citywide lockdown policy during this time frame \cite{birge2022controlling}, most of the workplaces were closed and only essential businesses like grocery stores were allowed to operate. Therefore, we contend that the most trips to grocery stores during that time frame could be considered single-purpose trips rather than multi-purpose ones \cite{lucchini2021living}, and thus we are able to use distance between the store and customer home location in the model. Moreover, due to lowered mobility within the lockdown period, we employ \emph{census tract} level visit patterns to capture broader mobility observations. We treat each census tract as an individual mobility center and generate a model on the aggregated mobility patterns within the specified time range so that the probabilities of residents within a census tract visiting given POIs are obtained.

We add randomly generated hypothetical grocery stores with area equal to the mean store area within Staten Island and continue adding until the new grocery per 1K residents match the grocery store densities of Manhattan and Queens, boroughs with more similar demographics. We use the existing mobility data to simulate visit numbers with respect to the previously obtained probabilities for the hypothetical setting and obtain the synthetic visit numbers for all POIs. We aggregate the visits in each census tract both for the ground truth and simulated cases, and then analyze the change in visits by residents of Staten Island to hotspot CBGs, that frequently appear in the top new cases quartile. The simulation results show that the visits to hotspot CBGs decreased by 47\% and 23\% by Manhattan and Queens' grocery store densities, respectively, which in turn means less exposure to possible COVID-19 spreaders and lower risk of contamination and reduced mortality (for more details see SI Notes 2).

\section*{Discussion}
In this research we build on the tradition of using network structures to explain human behavior in socioeconomic settings 
by examining the complex relationship between human mobility, socioeconomic outcomes, and demographic attributes within the context of the COVID-19 pandemic. In particular,  we take a network analysis approach to understanding the impact of the COVID-19 pandemic on the mobility patterns of the residents of NYC within its five boroughs and 6,493 census block groups (CBGs) during 2020. The network nodes represent CBGs, and the links correspond to visits between pairs of CBGs using visits to POIs by residents. We investigate node-specific and ego-network based structural features to compute dissimilarity scores between weekly networks year-on-year for 2019 and 2020 to understand the magnitude of change in the network structure. 

In addition, we analyze the temporal changes in node and degree centrality measurements across different socioeconomic groups. Our approach and findings reveal that while the COVID-19 response measures resulted in structural changes in mobility network, the CBGs that changed their ego-network structure the least had higher COVID-19 infection rates. The majority of such nodes are from low income and low education level neighborhoods with higher rates of front-line workers \cite{nycfrontline} (e.g., workers in healthcare, grocery, convenience and drug stores, child care, food and family services, public transport, trucking, warehouse and postal services) who could not reduce their mobility by as much. Arguably this is because of their job type that requires working outside their home and commuting frequently.

Our CBG-level dissimilarity analysis suggests a clear demographic distinction of residents who live in top-dissimilarity CBGs that appeared in more than 60\% of the weekly patterns we analyzed: high-income, higher education level, and mostly white population. We argue that these are residents whose job types can most suitably respond to shelter-in-place and physical/social distancing orders through remote work, unlike the residents in neighborhoods with other job types. Hence, in line with other studies \cite{chetty2020economic, chang2021mobility, hunter2021effect} we find that these top-dissimilarity CBGs and their residents are more resilient under the conditions of a pandemic such as COVID-19.

\subsubsection*{Two-sides of the same city}
Using a network perspective, we extend the previous work by finding that the adaptability of mobility patterns varies significantly not only by socioeconomic and demographic features but also by geographical attributes of neighborhoods. Quantifying the ego-network dissimilarity, we show that the less affluent and less educated neighborhoods had less adaptability to policy interventions aimed at reducing their mobility level. In addition, we find that the neighborhoods with higher income and education can show a similar behavior to less affluent and less educated neighborhoods if they have relatively limited access to public transport, workplaces, shops, and a set of other diverse amenities, as is the case with Staten Island. Even though there exist very few physical connections between Staten Island and other boroughs of NYC and New Jersey, which could conceivably mean that the existing mobility network structure would be expected to break down when there is an extreme exogenous shock, in contrast we find that the network changes were the opposite: They were very minimal. Thus, Staten Island was relatively fragile to the COVID-19 pandemic in terms of infections. Exploring similar neighborhoods or isolated geographical blocks in other urban areas could be the subject of future research to help policy makers develop effective policies to alleviate the burden of a pandemic in such areas and increase their resilience.

\subsubsection*{Opportunities for urban planning}
Our study provides valuable insights for urban planning and policy: Both socioeconomic as well as geographic characteristics of neighborhoods (the physical structures of neighborhoods and communities in an urban area) are important considerations when trying to increase neighborhoods' resilience to future exogenous shocks. With regards to the latter and based on our simulation results, convenient access to POIs that provide daily essential needs (e.g. grocery stores), workplaces, and centers of attraction that offer a variety of amenities through a diverse set of POIs would be expected to reduce the need for traveling long distances.
Using a Huff Gravity Model we performed a hypothetical scenario analysis to estimate the number of visits to COVID-19 hotspot areas with hypothetically added grocery stores for each CBG in Staten Island. Although we only consider customer-store distance and POI floor area in our model, the results yield valuable insights on the simulated mobility. Our findings indicate that as a result of increase in access to the essential POIs, residents are less exposed to hotspot districts and are able to satisfy their needs without traveling further distances. In cases like the COVID-19 pandemic, increased level of access to POIs could potentially decrease the level of infections and save more lives.
As the continued pandemic conditions reveal, these factors might be a major influence on future work habits and trends, and resilient communities organized and structured along these lines might be the preferred choice for many members of the population. As a future research direction, we aim to extend our simulation analysis by incorporating more POI categories with additional POI attraction features such as a neighborhood's physical characteristic into the model to evaluate the results within different settings.

\subsubsection*{Limitations}
Our study has limitations that arise mainly from the datasets used. The mobility patterns and visits to POIs are aggregated by time (e.g., weekly) and neighborhood (e.g., CBGs). While this method preserves the privacy of smartphone users, we were not able, as a result, to capture the primary aim of users' trips to different destination CBGs (e.g., for work, shopping, entertainment or other purposes). This additional level of detail would allow us to distinguish between essential and non-essential trips made by residents and match it with socioeconomic characteristics of neighborhoods.
Originating from the same limitation, we are only able to see which neighborhoods are connected to one another in the form of visits and with what frequency, but not how (i.e. mode of transportation) they are connected. While it might be possible to infer mostly-used transportation modes and routes between CBGs using other datasets, the resolution of our data does not allow us to link it with individual trips, preventing us from conducting additional analysis on what portion of trips are made in what modes and its relations with the spreading of the virus and the resilience of communities.

Furthermore, the POI dataset we use in our study mostly includes places where financial transactions take place (e.g. supermarkets, dentist offices, restaurants, etc.) and has a low coverage of workplaces and offices (e.g. corporate buildings, co-work spaces, etc.). The latter would, again, help us identify additional types of essential travel (e.g., workplaces) and better fit the gravity model for simulation purposes. We tried to address this limitation partially by using complementary datasets such as Google's mobility report and the mobility survey results published by NYC officials.

Finally, despite all the limitations, our study makes at least two contributions. First, our study results contribute substantive insights into the heterogeneity of mobility patterns among different neighborhoods during the COVID-19 pandemic through leveraging network science approaches and quantitative scenario analysis. Second, our study illuminates the factors associated with a neighborhood’s resilience and fragility. Understanding the factors associated with a neighborhood's resilience and fragility can help urban planners and authorities recommend sustainable policies, make better intervention decisions and be prepared to react more effectively to future exogenous shocks like the COVID-19 pandemic and save more lives.

\section*{Methods} \label{methods}

    \subsection*{Datasets}

        \subsubsection*{SafeGraph Mobility and Places Data}
        
            The dataset released by \emph{SafeGraph} (\url{www.safegraph.com}) provides fine-grained user geo-location information collected through the smartphone applications of millions of users who consented to share their live location. The data is anonymized and then aggregated by the total visits from CBGs to POIs of various categories (e.g. restaurants, grocery stores, and supermarkets) within a defined time period (i.e. weekly and monthly). More specifically, we use the \emph{Weekly Patterns} \cite{safegraph} dataset that provides the weekly aggregated visits from January 2019 to December 2020. The patterns dataset is geographically filtered with respect to the administrative boundaries of the New York Metropolitan Area, which resulted in 6,463 CBGs (in New York City) and 333,241 POIs in total.
        
        \subsubsection*{Google COVID-19 Community Mobility Reports}
            This dataset is prepared by Google to provide insights into mobility trends in order to help understand communities' response to interventions against COVID-19 pandemic. The data is collected from various smartphone applications including Google Maps. The data shows dynamics of mobility trends by region, across different POI categories compared to a baseline day that represents a normal day before the pandemic. The baseline is the median value from the five‑week period from January 3rd to February 6th, 2020. The POI categories include retail and recreation, groceries and pharmacies, parks, transit stations, workplaces, and residential buildings. 
        
        \subsubsection*{COVID-19 Cases}
            The Center for Systems Science and Engineering (CSSE) at Johns Hopkins University \cite{Badr2021.05.05.21256712} provides a COVID-19 data catalog presenting the latest updates all around the world with differing granularity levels. In NYC, the statistics for variables such as new cases, test counts, and deaths, are provided by ZIP Code Tabulation Areas (ZCTAs). A CBG may be located inside the borders of multiple ZCTAs. In order to work with the estimated weekly cases per CBG, the numbers were inferred by taking a weighted average based on the ratio of population of a CBG in each ZCTA and its corresponding COVID-19 case rate. Assume $S$ is the set of ZCTAs that geographically intersect with $CBG_i$. Then the number of COVID-19 cases for $CBG_i$ at time $t$ are estimated using the following formulations.
            
            \begin{ceqn}
                \begin{equation}
                    w_{ij} = \frac{population\; of\;\;   CBG_i\;\;   in\;\;   ZCTA_j}{population\;\;   of\;\;   ZCTA_j}
                \end{equation}
            \end{ceqn}
            
            \begin{ceqn}
                \begin{equation}
                    CCBG^t_i = \sum_{j \in S} (w_{ij} \times case^t_j)
                \end{equation}
            \end{ceqn}
            
            $CCBG^t_i$ denotes the estimated COVID-19 cases in $CBG_i$ at time $t$, and $case^t_j$ is the number of COVID cases in $ZCTA_j$ at time $t$.

        \subsubsection*{The United States Census Data}
        
            The American Community Survey (ACS) of the U.S. Census Bureau reports the estimates of demographic features at a CBG level. The most recently published 5-year ACS data collected in 2019, is utilized to extract the demographic features, namely: total population, median household income, education level, commuting time, and population distribution by race for each CBG. The resulting features are also represented with corresponding percentile levels.
            
        \subsubsection*{New York Metropolitan Area}
            In 2019, New York Metropolitan Area hosted the highest population at 19.22 million as one of the leading economies in the US with 1.522 Billion Dollars\cite{ny-statista}. The metropolitan area consists of counties from four different states, New York,  New Jersey,  Connecticut, and  Pennsylvania. In total, 7,809 census tracts and 23 counties are represented. POIs provided by SafeGraph Patterns are filtered by their location with respect to the New York Metropolitan area. 
            New York City (NYC) with about 9 million residents is the largest city in the New York Metropolitan area containing 6,493 CBGs. NYC is also divided into five administrative boundaries called boroughs. Those five boroughs are: Manhattan, Brooklyn, Bronx, Queens, and Staten Island.

    \subsection*{Mobility Networks}
        We model the mobility patterns among CBGs as weighted directed networks, $G^{(t)} \, = \, (V^{(t)}, E^{(t)})$, for each time step $t$, in which the nodes $V^{(t)}$ correspond to CBGs and edges $E^{(t)}$ capture the weekly number of visits initiating from a CBG to another with weight $w_{ij}^{(t)}$ representing the number of visits from $CBG_i$ to $CBG_j$ in time step $t$. Time steps are weeks from January 2019 and December 2020. In the \emph{Weekly Patterns} dataset, the aggregated number of visits from CBGs to a particular POI is stored. Here, the CBG where the destination POI is located, is defined as the target CBG. This means the POIs $P \, = \, \{p_1, ..., p_n\}$ inside $CBG_i$ are used as a proxy to obtain the in-flow mobility, and their aggregated number of visitors constitutes the weights of the incoming edges for $CBG_i$. Furthermore, the road network distances between CBGs are also added as edge attributes into the network metrics.

    \subsection*{Topological changes over time}
    
        In order to display the temporal changes in the network topology, we first focus on the dissimilarity analysis between paired weekly networks of 2019 and 2020 and further break the nodes down based on demographic groups. Next, we analyze the relationship between centrality metrics and demographic groups.
    
        \subsubsection*{Dissimilarity analysis}
        
            To form the feature vectors of the aligned weekly networks of 2019 and 2020, we consider the following ego-network based node features as proposed by Berlingerio et al. \cite{netsimile}.
            
            \begin{itemize}
               \item Node In \& Out Degree
               \item Node Strength
               \item Clustering Coefficient
               \item Average Degree \& Average Clustering Coefficient of Node's Neighborhood
               \item Number of Edges \& Alters in Node's Ego-Network
               \item Number of Out-going Edges from Node's Ego-Network
             \end{itemize}
            
            To compute the overall dissimilarity score between the paired weekly networks, the resulting node feature vectors are merged into a single vector, in which the network features are obtained by applying statistical aggregations, such as standard deviation, median, kurtosis, and skewness on the node features. Next, in the node level dissimilarity analysis, paired node feature vectors are utilized without any aggregation and a score is generated for each node. For both analyses, Canberra distance \cite{lance1967mixed} is applied on the paired feature vectors $P$ and $Q$ using Equation \ref{canberra distance}.
            
            \begin{ceqn}
                \begin{equation} \label{canberra distance}
                   d(P, Q) = \sum_{i=1} \, \frac{|P \, - \, Q|}{|P| + |Q|}
                \end{equation}
            \end{ceqn}
            
        \subsubsection*{Centrality Evolution}
        
            Neighborhoods with different demographic backgrounds manifest varying repercussions to the enforced NPIs over time, due to their distinctive needs and socioeconomic dynamics. In parallel, CBGs with ample POIs became frequent destinations. To this end, centrality metrics are employed to demonstrate the temporal change in the topological importance of the CBGs. In particular, we focus on \emph{betweenness}, \emph{in-degree}, \emph{out-degree} and \emph{self-visit ratio}. In order to estimate the number of visits to the POIs in the home CBG, the self-visit ratio is defined as:
            
            \begin{ceqn}
                \begin{equation} \label{self-visit-eq}
                    S_c^{t} \, = \, \frac{W_l^t}{W_l^t \, + \, W_o^t}
                \end{equation}
            \end{ceqn}
            
        In Equation \ref{self-visit-eq} $W_l^t$ is the sum of weights on self-loops and $W_o^t$ is the sum of weights on outgoing edges in time step $t$ for CBG $c$. The self-visit ratio is considered as an indicator of locality of residents visits.

            
    \subsection*{COVID-19 Hotspots and Bridge CBGs}
    
        The mobility per CBG is a function of multiple parameters which are heavily affected by intricate social dynamics. Because of the evident correlation between mobility and spreading of COVID-19 infections \cite{chang2021mobility, iacus2020human, xiong2020mobile}, CBGs with higher mobility rates may cause a potential threat to restraining the disease prevalence. 
        We refer to the CBGs that rank in the top weekly new COVID-19 cases as COVID-19 \textit{hotspots}, and the CBGs that interact frequently with the hotspots as COVID-19 \textit{bridge} CBGs. 
        To detect and analyze the potential spreaders or bridges, we utilize the visit frequencies between CBGs. First, we consider two weeks long time span as the incubation period for new cases to emerge \cite{chang2021mobility, iacus2020human}. Starting from March 2nd, 2020, when the first case in NYC was reported, we isolate the CBGs in the top weekly new cases quartile in week $t$ (hotspots). Next, the CBGs that have an outgoing edge to the isolated group in week $t-2$ are recorded. To determine the COVID-19 bridge CBGs, a percentile based frequency filtering is applied. To this end, we investigate CBGs with occurrence frequencies in the $75^{th}$ percentile as possible bridges. The aim of this analysis is to find out if the bridge CBGs have special socioeconomic and demographic features and to understand why they show such mobility patterns.
        
    \subsection*{Huff Gravity Model}
    
        The Huff gravity model \cite{huff1964defining} is a well-known and widely used market share estimation model that focuses on the relationship between retail stores and customers by modelling it as a function of distance and store attractiveness traits such as store area. Equation \ref{huff} formalizes the basic version of the gravity model that we use in our analyses.
        
        \begin{ceqn}
            \begin{equation} \label{huff}
                P_{ij} = \frac{\frac{A_{j}^\alpha}{D_{ij}^\beta}}{\sum_{k=1}^n \frac{A_{k}^\alpha}{D_{ik}^\beta}}
            \end{equation}
        \end{ceqn}
        
        $P_{ij}$ corresponds to the probability of customer $i$ choosing POI $j$ for shopping among all available stores in her choice set $k$. The exponents $\alpha$ and $\beta$ control the weight of distance and store area. Since there was a significant decline in movement activities during the COVID-19 pandemic, in order to have reasonable amount of observations that can provide a better fit for the model, we aggregate the visits from CBGs into a census tract level. We consider each census tract as an individual mobility center and estimate a pair of parameters $\alpha$ and $\beta$ for each census tract using Particle Swarm Optimization technique\cite{suhara2021validating, liang2020calibrating, bahrami2022using}.
        
    \subsection*{Availability of Data and Code}
    
        All the datasets and scripts used for this study are freely available for further research and replication purposes. \\
        The Safegraph mobility dataset is freely available for academic research purposes through request at: \\ \url{https://www.safegraph.com/academics}\\
        The Google COVID-19 Community Mobility Reports data is available at:\\ \url{https://www.google.com/covid19/mobility/}\\
        All scripts are available at the project github repository:\\ \url{https://github.com/alppboz/safegraph-covid19-mobility}

    \subsection*{Abbreviations}
        CBG: Census Block Group\\
        NPI: Non-pharmaceutical Intervention\\
        NYC: New York City\\
        POI: Point of Interest
        

\bibliography{main}

\begin{thebibliography}{10}
\urlstyle{rm}
\expandafter\ifx\csname url\endcsname\relax
  \def\url#1{\texttt{#1}}\fi
\expandafter\ifx\csname urlprefix\endcsname\relax\def\urlprefix{URL }\fi
\expandafter\ifx\csname doiprefix\endcsname\relax\def\doiprefix{DOI: }\fi
\providecommand{\bibinfo}[2]{#2}
\providecommand{\eprint}[2][]{\url{#2}}

\bibitem{chong2020economic}
\bibinfo{author}{Chong, S.~K.} \emph{et~al.}
\newblock \bibinfo{journal}{\bibinfo{title}{Economic outcomes predicted by
  diversity in cities}}.
\newblock {\emph{\JournalTitle{EPJ Data Science}}}
  \textbf{\bibinfo{volume}{9}}, \bibinfo{pages}{17} (\bibinfo{year}{2020}).

\bibitem{singh2015money}
\bibinfo{author}{Singh, V.~K.}, \bibinfo{author}{Bozkaya, B.} \&
  \bibinfo{author}{Pentland, A.}
\newblock \bibinfo{journal}{\bibinfo{title}{Money walks: implicit mobility
  behavior and financial well-being}}.
\newblock {\emph{\JournalTitle{PloS one}}} \textbf{\bibinfo{volume}{10}},
  \bibinfo{pages}{e0136628} (\bibinfo{year}{2015}).

\bibitem{bettencourt2013origins}
\bibinfo{author}{Bettencourt, L.~M.}
\newblock \bibinfo{journal}{\bibinfo{title}{The origins of scaling in cities}}.
\newblock {\emph{\JournalTitle{Science}}} \textbf{\bibinfo{volume}{340}},
  \bibinfo{pages}{1438--1441} (\bibinfo{year}{2013}).

\bibitem{buera2020global}
\bibinfo{author}{Buera, F.~J.} \& \bibinfo{author}{Oberfield, E.}
\newblock \bibinfo{journal}{\bibinfo{title}{The global diffusion of ideas}}.
\newblock {\emph{\JournalTitle{Econometrica}}} \textbf{\bibinfo{volume}{88}},
  \bibinfo{pages}{83--114} (\bibinfo{year}{2020}).

\bibitem{sveikauskas1975productivity}
\bibinfo{author}{Sveikauskas, L.}
\newblock \bibinfo{journal}{\bibinfo{title}{The productivity of cities}}.
\newblock {\emph{\JournalTitle{The Quarterly Journal of Economics}}}
  \textbf{\bibinfo{volume}{89}}, \bibinfo{pages}{393--413}
  (\bibinfo{year}{1975}).

\bibitem{bettencourt2007growth}
\bibinfo{author}{Bettencourt, L.~M.}, \bibinfo{author}{Lobo, J.},
  \bibinfo{author}{Helbing, D.}, \bibinfo{author}{K{\"u}hnert, C.} \&
  \bibinfo{author}{West, G.~B.}
\newblock \bibinfo{journal}{\bibinfo{title}{Growth, innovation, scaling, and
  the pace of life in cities}}.
\newblock {\emph{\JournalTitle{Proceedings of the National Academy of
  Sciences}}} \textbf{\bibinfo{volume}{104}}, \bibinfo{pages}{7301--7306}
  (\bibinfo{year}{2007}).

\bibitem{alvarez2013idea}
\bibinfo{author}{Alvarez, F.~E.}, \bibinfo{author}{Buera, F.~J.},
  \bibinfo{author}{Lucas, R.~E.} \emph{et~al.}
\newblock \bibinfo{title}{Idea flows, economic growth, and trade}.
\newblock \bibinfo{type}{Tech. Rep.}, \bibinfo{institution}{National Bureau of
  Economic Research} (\bibinfo{year}{2013}).

\bibitem{steele2017mapping}
\bibinfo{author}{Steele, J.~E.} \emph{et~al.}
\newblock \bibinfo{journal}{\bibinfo{title}{Mapping poverty using mobile phone
  and satellite data}}.
\newblock {\emph{\JournalTitle{Journal of The Royal Society Interface}}}
  \textbf{\bibinfo{volume}{14}}, \bibinfo{pages}{20160690}
  (\bibinfo{year}{2017}).

\bibitem{pan2013urban}
\bibinfo{author}{Pan, W.}, \bibinfo{author}{Ghoshal, G.},
  \bibinfo{author}{Krumme, C.}, \bibinfo{author}{Cebrian, M.} \&
  \bibinfo{author}{Pentland, A.}
\newblock \bibinfo{journal}{\bibinfo{title}{Urban characteristics attributable
  to density-driven tie formation}}.
\newblock {\emph{\JournalTitle{Nature Communications}}}
  \textbf{\bibinfo{volume}{4}}, \bibinfo{pages}{1--7} (\bibinfo{year}{2013}).

\bibitem{schlapfer2021universal}
\bibinfo{author}{Schl{\"a}pfer, M.} \emph{et~al.}
\newblock \bibinfo{journal}{\bibinfo{title}{The universal visitation law of
  human mobility}}.
\newblock {\emph{\JournalTitle{Nature}}} \textbf{\bibinfo{volume}{593}},
  \bibinfo{pages}{522--527} (\bibinfo{year}{2021}).

\bibitem{lazer2009life}
\bibinfo{author}{Lazer, D.} \emph{et~al.}
\newblock \bibinfo{journal}{\bibinfo{title}{Life in the network: the coming age
  of computational social science}}.
\newblock {\emph{\JournalTitle{Science (New York, NY)}}}
  \textbf{\bibinfo{volume}{323}}, \bibinfo{pages}{721} (\bibinfo{year}{2009}).

\bibitem{flaxman2020estimating}
\bibinfo{author}{Flaxman, S.} \emph{et~al.}
\newblock \bibinfo{journal}{\bibinfo{title}{Estimating the effects of
  non-pharmaceutical interventions on \uppercase{COVID}-19 in
  \uppercase{E}urope}}.
\newblock {\emph{\JournalTitle{Nature}}} \textbf{\bibinfo{volume}{584}},
  \bibinfo{pages}{257--261} (\bibinfo{year}{2020}).

\bibitem{kraemer2020effect}
\bibinfo{author}{Kraemer, M.~U.} \emph{et~al.}
\newblock \bibinfo{journal}{\bibinfo{title}{The effect of human mobility and
  control measures on the covid-19 epidemic in \uppercase{C}hina}}.
\newblock {\emph{\JournalTitle{Science}}} \textbf{\bibinfo{volume}{368}},
  \bibinfo{pages}{493--497} (\bibinfo{year}{2020}).

\bibitem{berry2021evaluating}
\bibinfo{author}{Berry, C.~R.}, \bibinfo{author}{Fowler, A.},
  \bibinfo{author}{Glazer, T.}, \bibinfo{author}{Handel-Meyer, S.} \&
  \bibinfo{author}{MacMillen, A.}
\newblock \bibinfo{journal}{\bibinfo{title}{Evaluating the effects of
  shelter-in-place policies during the covid-19 pandemic}}.
\newblock {\emph{\JournalTitle{Proceedings of the National Academy of
  Sciences}}} \textbf{\bibinfo{volume}{118}} (\bibinfo{year}{2021}).

\bibitem{aleta2020modelling}
\bibinfo{author}{Aleta, A.} \emph{et~al.}
\newblock \bibinfo{journal}{\bibinfo{title}{Modelling the impact of testing,
  contact tracing and household quarantine on second waves of
  \uppercase{COVID}-19}}.
\newblock {\emph{\JournalTitle{Nature Human Behaviour}}}
  \textbf{\bibinfo{volume}{4}}, \bibinfo{pages}{964--971}
  (\bibinfo{year}{2020}).

\bibitem{schlosser2020covid}
\bibinfo{author}{Schlosser, F.} \emph{et~al.}
\newblock \bibinfo{journal}{\bibinfo{title}{Covid-19 lockdown induces
  disease-mitigating structural changes in mobility networks}}.
\newblock {\emph{\JournalTitle{Proceedings of the National Academy of
  Sciences}}} \textbf{\bibinfo{volume}{117}}, \bibinfo{pages}{32883--32890}
  (\bibinfo{year}{2020}).

\bibitem{chetty2020economic}
\bibinfo{author}{Chetty, R.}, \bibinfo{author}{Friedman, J.~N.},
  \bibinfo{author}{Hendren, N.}, \bibinfo{author}{Stepner, M.} \&
  \bibinfo{author}{Team, T. O.~I.}
\newblock \emph{\bibinfo{title}{The economic impacts of COVID-19: Evidence from
  a new public database built using private sector data}}.
\newblock \bibinfo{number}{w27431} (\bibinfo{publisher}{National Bureau of
  Economic Research}, \bibinfo{year}{2020}).

\bibitem{gao2020mapping}
\bibinfo{author}{Gao, S.}, \bibinfo{author}{Rao, J.}, \bibinfo{author}{Kang,
  Y.}, \bibinfo{author}{Liang, Y.} \& \bibinfo{author}{Kruse, J.}
\newblock \bibinfo{journal}{\bibinfo{title}{Mapping county-level mobility
  pattern changes in the united states in response to \uppercase{COVID}-19}}.
\newblock {\emph{\JournalTitle{SIGSpatial Special}}}
  \textbf{\bibinfo{volume}{12}}, \bibinfo{pages}{16--26}
  (\bibinfo{year}{2020}).

\bibitem{galeazzi2020human}
\bibinfo{author}{Galeazzi, A.} \emph{et~al.}
\newblock \bibinfo{journal}{\bibinfo{title}{Human mobility in response to
  covid-19 in \uppercase{F}rance, \uppercase{I}taly and \uppercase{UK}}}.
\newblock {\emph{\JournalTitle{arXiv preprint arXiv:2005.06341}}}
  (\bibinfo{year}{2020}).

\bibitem{chang2021mobility}
\bibinfo{author}{Chang, S.} \emph{et~al.}
\newblock \bibinfo{journal}{\bibinfo{title}{Mobility network models of covid-19
  explain inequities and inform reopening}}.
\newblock {\emph{\JournalTitle{Nature}}} \textbf{\bibinfo{volume}{589}},
  \bibinfo{pages}{82--87} (\bibinfo{year}{2021}).

\bibitem{netsimile}
\bibinfo{author}{Berlingerio, M.}, \bibinfo{author}{Koutra, D.},
  \bibinfo{author}{Eliassi-Rad, T.} \& \bibinfo{author}{Faloutsos, C.}
\newblock \bibinfo{journal}{\bibinfo{title}{Netsimile: A scalable approach to
  size-independent network similarity}}.
\newblock {\emph{\JournalTitle{arXiv preprint arXiv:1209.2684}}}
  (\bibinfo{year}{2012}).

\bibitem{hunter2021effect}
\bibinfo{author}{Hunter, R.~F.} \emph{et~al.}
\newblock \bibinfo{journal}{\bibinfo{title}{Effect of covid-19 response
  policies on walking behavior in us cities}}.
\newblock {\emph{\JournalTitle{Nature Communications}}}
  \textbf{\bibinfo{volume}{12}}, \bibinfo{pages}{1--9} (\bibinfo{year}{2021}).

\bibitem{heroy2021covid}
\bibinfo{author}{Heroy, S.}, \bibinfo{author}{Loaiza, I.},
  \bibinfo{author}{Pentland, A.} \& \bibinfo{author}{O’Clery, N.}
\newblock \bibinfo{journal}{\bibinfo{title}{Covid-19 policy analysis: labour
  structure dictates lockdown mobility behaviour}}.
\newblock {\emph{\JournalTitle{Journal of the Royal Society Interface}}}
  \textbf{\bibinfo{volume}{18}}, \bibinfo{pages}{20201035}
  (\bibinfo{year}{2021}).

\bibitem{deville2016scaling}
\bibinfo{author}{Deville, P.} \emph{et~al.}
\newblock \bibinfo{journal}{\bibinfo{title}{Scaling identity connects human
  mobility and social interactions}}.
\newblock {\emph{\JournalTitle{Proceedings of the National Academy of
  Sciences}}} \textbf{\bibinfo{volume}{113}}, \bibinfo{pages}{7047--7052}
  (\bibinfo{year}{2016}).

\bibitem{miritello2011dynamical}
\bibinfo{author}{Miritello, G.}, \bibinfo{author}{Moro, E.} \&
  \bibinfo{author}{Lara, R.}
\newblock \bibinfo{journal}{\bibinfo{title}{Dynamical strength of social ties
  in information spreading}}.
\newblock {\emph{\JournalTitle{Physical Review E}}}
  \textbf{\bibinfo{volume}{83}}, \bibinfo{pages}{045102}
  (\bibinfo{year}{2011}).

\bibitem{googlemobility}
\bibinfo{title}{{Google COVID-19 Community Mobility Reports}}.
\newblock
  \bibinfo{howpublished}{\url{https://www.google.com/covid19/mobility/}}.
\newblock \bibinfo{note}{Accessed: 2022-09-29}.

\bibitem{nycinsouts}
\bibinfo{title}{{The Ins and Outs of NYC Commuting}}.
\newblock
  \bibinfo{howpublished}{\url{https://www1.nyc.gov/assets/planning/download/pdf/planning-level/housing-economy/nyc-ins-and-out-of-commuting.pdf}}.
\newblock \bibinfo{note}{Accessed: 2022-09-29}.

\bibitem{nycmobilityreport}
\bibinfo{title}{{New York City Mobility Report 2019}}.
\newblock
  \bibinfo{howpublished}{\url{http://www.nyc.gov/html/dot/downloads/pdf/mobility-report-2019-print.pdf}}.
\newblock \bibinfo{note}{Accessed: 2022-09-29}.

\bibitem{huff1964defining}
\bibinfo{author}{Huff, D.~L.}
\newblock \bibinfo{journal}{\bibinfo{title}{Defining and estimating a trading
  area}}.
\newblock {\emph{\JournalTitle{Journal of Marketing}}}
  \textbf{\bibinfo{volume}{28}}, \bibinfo{pages}{34--38}
  (\bibinfo{year}{1964}).

\bibitem{birge2022controlling}
\bibinfo{author}{Birge, J.~R.}, \bibinfo{author}{Candogan, O.} \&
  \bibinfo{author}{Feng, Y.}
\newblock \bibinfo{journal}{\bibinfo{title}{Controlling epidemic spread:
  Reducing economic losses with targeted closures}}.
\newblock {\emph{\JournalTitle{Management Science}}}  (\bibinfo{year}{2022}).

\bibitem{lucchini2021living}
\bibinfo{author}{Lucchini, L.} \emph{et~al.}
\newblock \bibinfo{journal}{\bibinfo{title}{Living in a pandemic: changes in
  mobility routines, social activity and adherence to covid-19 protective
  measures}}.
\newblock {\emph{\JournalTitle{Scientific reports}}}
  \textbf{\bibinfo{volume}{11}}, \bibinfo{pages}{1--12} (\bibinfo{year}{2021}).

\bibitem{nycfrontline}
\bibinfo{title}{{New York City’s Frontline Workers}}.
\newblock
  \bibinfo{howpublished}{\url{https://comptroller.nyc.gov/reports/new-york-citys-frontline-workers/\#how-our-frontline-workers-get-to-work}}.
\newblock \bibinfo{note}{Accessed: 2022-09-29}.

\bibitem{safegraph}
\bibinfo{author}{SafeGraph}.
\newblock \bibinfo{title}{\uppercase{W}eekly \uppercase{P}atterns}.
\newblock \bibinfo{howpublished}{Available at
  \url{https://docs.safegraph.com/docs/weekly-patterns} (2021)}.

\bibitem{Badr2021.05.05.21256712}
\bibinfo{author}{Badr, H.~S.} \emph{et~al.}
\newblock \bibinfo{journal}{\bibinfo{title}{Unified real-time
  environmental-epidemiological data for multiscale modeling of the
  \uppercase{COVID}-19 pandemic}}.
\newblock {\emph{\JournalTitle{medRxiv}}}
  \doiprefix\url{10.1101/2021.05.05.21256712} (\bibinfo{year}{2021}).
\newblock
  \eprint{https://www.medrxiv.org/content/early/2021/05/07/2021.05.05.21256712.full.pdf}.

\bibitem{ny-statista}
\bibinfo{author}{Statista}.
\newblock \bibinfo{title}{Gdp of the new york metro area from 2001 to 2020}.
\newblock \bibinfo{howpublished}{Available at
  \url{https://www.statista.com/statistics/183815/gdp-of-the-new-york-metro-area/}
  (2021)}.

\bibitem{lance1967mixed}
\bibinfo{author}{Lance, G.~N.} \& \bibinfo{author}{Williams, W.~T.}
\newblock \bibinfo{journal}{\bibinfo{title}{Mixed-data classificatory programs
  i - agglomerative systems}}.
\newblock {\emph{\JournalTitle{Australian Computer Journal}}}
  \textbf{\bibinfo{volume}{1}}, \bibinfo{pages}{15--20} (\bibinfo{year}{1967}).

\bibitem{iacus2020human}
\bibinfo{author}{Iacus, S.~M.} \emph{et~al.}
\newblock \bibinfo{journal}{\bibinfo{title}{Human mobility and covid-19 initial
  dynamics}}.
\newblock {\emph{\JournalTitle{Nonlinear Dynamics}}}
  \textbf{\bibinfo{volume}{101}}, \bibinfo{pages}{1901--1919}
  (\bibinfo{year}{2020}).

\bibitem{xiong2020mobile}
\bibinfo{author}{Xiong, C.}, \bibinfo{author}{Hu, S.}, \bibinfo{author}{Yang,
  M.}, \bibinfo{author}{Luo, W.} \& \bibinfo{author}{Zhang, L.}
\newblock \bibinfo{journal}{\bibinfo{title}{Mobile device data reveal the
  dynamics in a positive relationship between human mobility and covid-19
  infections}}.
\newblock {\emph{\JournalTitle{Proceedings of the National Academy of
  Sciences}}} \textbf{\bibinfo{volume}{117}}, \bibinfo{pages}{27087--27089}
  (\bibinfo{year}{2020}).

\bibitem{suhara2021validating}
\bibinfo{author}{Suhara, Y.}, \bibinfo{author}{Bahrami, M.},
  \bibinfo{author}{Bozkaya, B.} \& \bibinfo{author}{Pentland, A.~S.}
\newblock \bibinfo{journal}{\bibinfo{title}{Validating gravity-based market
  share models using large-scale transactional data}}.
\newblock {\emph{\JournalTitle{Big Data}}} \textbf{\bibinfo{volume}{9}},
  \bibinfo{pages}{188--202} (\bibinfo{year}{2021}).

\bibitem{liang2020calibrating}
\bibinfo{author}{Liang, Y.}, \bibinfo{author}{Gao, S.}, \bibinfo{author}{Cai,
  Y.}, \bibinfo{author}{Foutz, N.~Z.} \& \bibinfo{author}{Wu, L.}
\newblock \bibinfo{journal}{\bibinfo{title}{Calibrating the dynamic
  \uppercase{H}uff model for business analysis using location big data}}.
\newblock {\emph{\JournalTitle{Transactions in GIS}}}
  \textbf{\bibinfo{volume}{24}}, \bibinfo{pages}{681--703}
  (\bibinfo{year}{2020}).

\bibitem{bahrami2022using}
\bibinfo{author}{Bahrami, M.}, \bibinfo{author}{Xu, Y.},
  \bibinfo{author}{Tweed, M.}, \bibinfo{author}{Bozkaya, B.} \&
  \bibinfo{author}{Pentland, A.~S.}
\newblock \bibinfo{journal}{\bibinfo{title}{Using gravity model to make store
  closing decisions: A data driven approach}}.
\newblock {\emph{\JournalTitle{Expert Systems with Applications}}}
  \bibinfo{pages}{117703} (\bibinfo{year}{2022}).

\end{thebibliography}


\begin{thebibliography}{1}

\bibitem{safegraph}
SafeGraph, ``Weekly patterns.'' Available at
  \url{https://docs.safegraph.com/docs/weekly-patterns} (2021).

\bibitem{suhara2021validating}
Y.~Suhara, M.~Bahrami, B.~Bozkaya, and A.~S. Pentland, ``Validating
  gravity-based market share models using large-scale transactional data,''
  {\em Big Data}, vol.~9, no.~3, pp.~188--202, 2021.

\bibitem{liang2020calibrating}
Y.~Liang, S.~Gao, Y.~Cai, N.~Z. Foutz, and L.~Wu, ``Calibrating the dynamic
  huff model for business analysis using location big data,'' {\em Transactions
  in GIS}, vol.~24, no.~3, pp.~681--703, 2020.

\bibitem{bahrami2022using}
M.~Bahrami, Y.~Xu, M.~Tweed, B.~Bozkaya, and A.~S. Pentland, ``Using gravity
  model to make store closing decisions: A data driven approach,'' {\em Expert
  Systems with Applications}, p.~117703, 2022.

\end{thebibliography}

\section*{Acknowledgements}
The authors would like to thank SafeGraph for making the mobility data available to this research project.

\section*{Author contributions}
Conceptualized the initial idea for the research: HAB, MB, NM, SB, BB, AP\\
Developed/designed research: HAB, MB, NM, SB, AN, BB\\
Prepared and analyzed the data: HAB, MB\\
Created figures/tables, and Github materials: HAB\\
Supervised the work: AP, SB\\
Wrote first draft: HAB, MB\\
Edited the draft: HAB, MB, NM, SB, AN, BB\\
All authors approved the final version.

\section*{Competing Interests}
The authors declare no competing interests.

\end{document}


\maketitle

\tableofcontents

\listoffigures
\listoftables



\section{Supplementary Note 1: Safegraph Weekly Patterns}

Safegraph \cite{safegraph} data consortium presents location aggregated weekly mobility data obtained from users who consented to anonymously share their GPS locations through mobile applications. In the dataset, weekly visits of the residents from \texttt{Census Block Groups} (CBG), a geographical unit with an approximate population of 600-3.000, to different \texttt{Point of Interests} (POI), such as grocery stores, restaurants and recreational hubs, are recorded. The dataset contains records from the USA and Canada starting from January 2019 and keeps getting updated periodically. In addition to the category of a POI, its location and CBG are also provided. We filter data points based on the location to retrieve all the records inside New York Metropolitan Area.

\begin{figure}[ht!]
\centering
\includegraphics[width=0.5\linewidth]{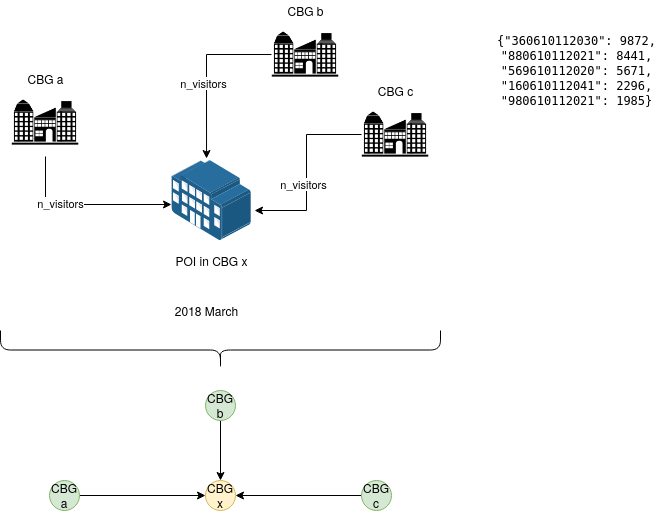}
\caption{CBG-CBG directed network creation from Safegraph patterns dataset. }
\label{fig:network_creation}
\end{figure}

In our study, we focus on the mobility between CBGs and try to understand the dynamics during the pandemic. In order to create such a network, we employed the latitude-longitude information of POIs to retrieve their corresponding CBG and accumulated the visits for that particular CBG. For each week, we created a directed network in which nodes consist of CBG and the edges between them capture the number of visitors. Figure \ref{fig:network_creation} depicts the procedure of creating CBG-CBG mobility networks in which visits from different CBGs to a POI in CBG \emph{x} are converted to an edge in the resulting networks.

\section{Supplementary Note 2: Tuning Huff Gravity Model}

Huff model evaluates the probability, $P_{ij}$, of residents of census tract \texttt{i} visiting the POI \texttt{j} with respect to the utility of POI \texttt{j} compared to its competitors as noted in Equation \ref{eq:huff-utility}. 

\begin{equation} \label{eq:huff-utility}
    P_{ij} = \frac{U_{ij}}{\sum_k U_{ik}}
\end{equation}

In this setting, utility consists of multiple factors such as store area and distance as we employed in our model, which yields Equation \ref{huff-model}.

\begin{equation} \label{huff-model}
    P_{ij} = \frac{\frac{A_{j}^\alpha}{D_{ij}^\beta}}{\sum_{k=1}^n \frac{A_{k}^\alpha}{D_{ik}^\beta}}
\end{equation}

\begin{figure}[ht!]
\centering
\includegraphics[width=0.4\linewidth]{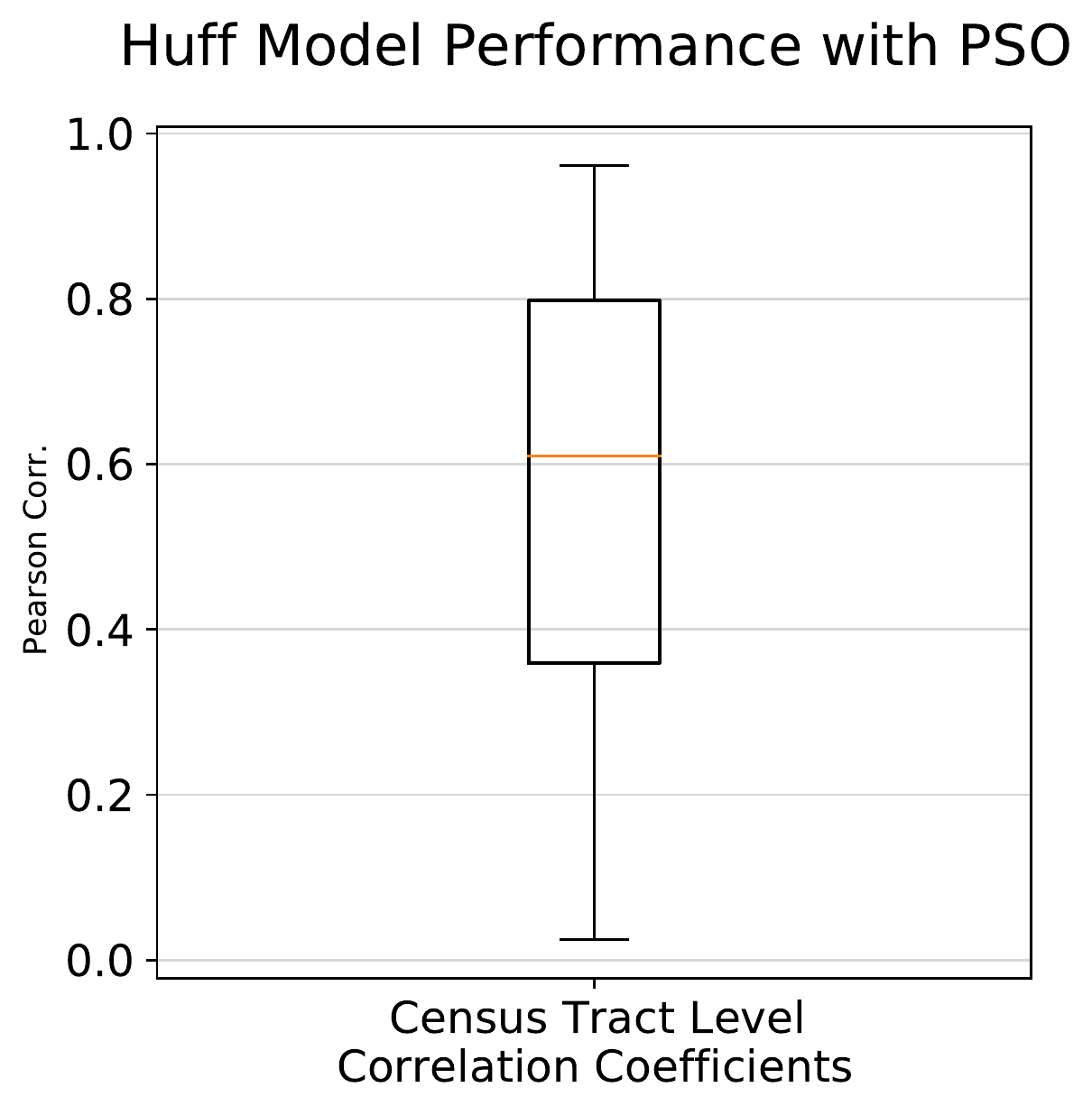}
\caption{Huff Model performance using parameters fitted by PSO.}
\label{fig:huff-performance}
\end{figure}

Since we model each census tract as an individual mobility center, we fit a model for each census tract (109 in total) and obtain $\alpha$ and $\beta$ exponents. In the literature, there exists multiple methods to estimate these exponents including Ordinary Least Squares, Geographically Weighted Regression, Generalized Linear Models, and Particle Swarm Optimization (PSO) techniques. Most recent studies show the PSO technique can provide better estimates for $\alpha$ and $\beta$ exponents \cite{suhara2021validating, liang2020calibrating, bahrami2022using}. Therefore, in this study we use PSO with objective function that maximized the Pearson Correlation between actual and estimated visit ratios. Equation \ref{pso-obj} displays the objective function in which the negated coefficient is minimized due to the software package we are using.

\begin{align} \label{pso-obj}
    a &= \text{actual visit ratios} \nonumber \\
    e_{\alpha, \beta}  &= \text{estimated visit ratios with respect to $\alpha$ \& $\beta$} \nonumber\\ 
    \argmin_{\alpha, \beta} &= 1 - pearson(a, e_{\alpha, \beta})
\end{align}

PSO is an iterative optimization method in which a number of particles collectively search for the best parameter set with in the search space. In our simulations, we searched within a 2-dimensional feature space (distance and store area) in 500 iterations with 20 particles. Cognitive and social parameters are both set to 1.5, while the inertia parameter is set to 0.9. Figure \ref{fig:huff-performance} displays the resulting Pearson Correlation coefficients for census tracts with a median of 0.61 showing good model fit.

\section{Supplementary Note 3: CBG Bridge \& Borough Regression Model}

To display the relationship between bridge occurrences and boroughs, we conduct a regression analysis in which dependent variable is ranked bridge occurrences for each CBG and independent variables are one-hot encoded boroughs. The results for the regression analysis can be found in Table \ref{table:reg}.

\begin{table}[!ht]
\begin{center}
\begin{tabular}{lclc}
\toprule
\textbf{Dep. Variable:}    &    Bridge Frequency Rank    & \textbf{  R-squared:         } &     0.201   \\
\textbf{Model:}            &       OLS        & \textbf{  Adj. R-squared:    } &     0.201   \\
\textbf{Method:}           &  Least Squares   & \textbf{  F-statistic:       } &     387.2   \\
\textbf{Date:}             & Mon, 19 Sep 2022 & \textbf{  Prob (F-statistic):} & 7.03e-298   \\
\textbf{Time:}             &     21:52:12     & \textbf{  Log-Likelihood:    } &   -393.46   \\
\textbf{No. Observations:} &        6146      & \textbf{  AIC:               } &     796.9   \\
\textbf{Df Residuals:}     &        6141      & \textbf{  BIC:               } &     830.5   \\
\textbf{Df Model:}         &           4      & \textbf{                     } &             \\
\bottomrule
\end{tabular}
\begin{tabular}{lcccccc}
                       & \textbf{coef} & \textbf{std err} & \textbf{t} & \textbf{P$> |$t$|$} & \textbf{[0.025} & \textbf{0.975]}  \\
\midrule
\textbf{Staten Island} &       0.8944  &        0.015     &    60.327  &         0.000        &        0.865    &        0.923     \\
\textbf{Manhattan}     &       0.3172  &        0.008     &    40.761  &         0.000        &        0.302    &        0.332     \\
\textbf{Bronx}         &       0.5699  &        0.008     &    72.737  &         0.000        &        0.555    &        0.585     \\
\textbf{Queens}        &       0.5701  &        0.006     &    89.566  &         0.000        &        0.558    &        0.583     \\
\textbf{Brooklyn}      &       0.4459  &        0.006     &    77.543  &         0.000        &        0.435    &        0.457     \\
\bottomrule
\end{tabular}
\begin{tabular}{lclc}
\textbf{Omnibus:}       & 463.528 & \textbf{  Durbin-Watson:     } &    0.355  \\
\textbf{Prob(Omnibus):} &   0.000 & \textbf{  Jarque-Bera (JB):  } &  166.915  \\
\textbf{Skew:}          &   0.123 & \textbf{  Prob(JB):          } & 5.69e-37  \\
\textbf{Kurtosis:}      &   2.231 & \textbf{  Cond. No.          } &     2.58  \\
\bottomrule
\end{tabular}
\caption{Bridge occurrences' regression model results with respect to boroughs.}
\label{table:reg}
\end{center}
\end{table}

\section{Supplementary Note 4: List of software used}

Analyses were conducted in Python programming language with following libraries:

\begin{itemize}

\item \texttt{Pandas} for I/O and data manipulations.

\item \texttt{GeoPandas} for spatial operations.

\item \texttt{Matplotlib} for visualizations.

\item \texttt{statsmodels} for statistical analyses.

\item \texttt{pyswarms} for Particle Swarm Optimization.

\end{itemize}

\section{Supplementary Note 5: Additional Figures}

\begin{figure}[hbt!]
\centering
\includegraphics[width=0.95\linewidth]{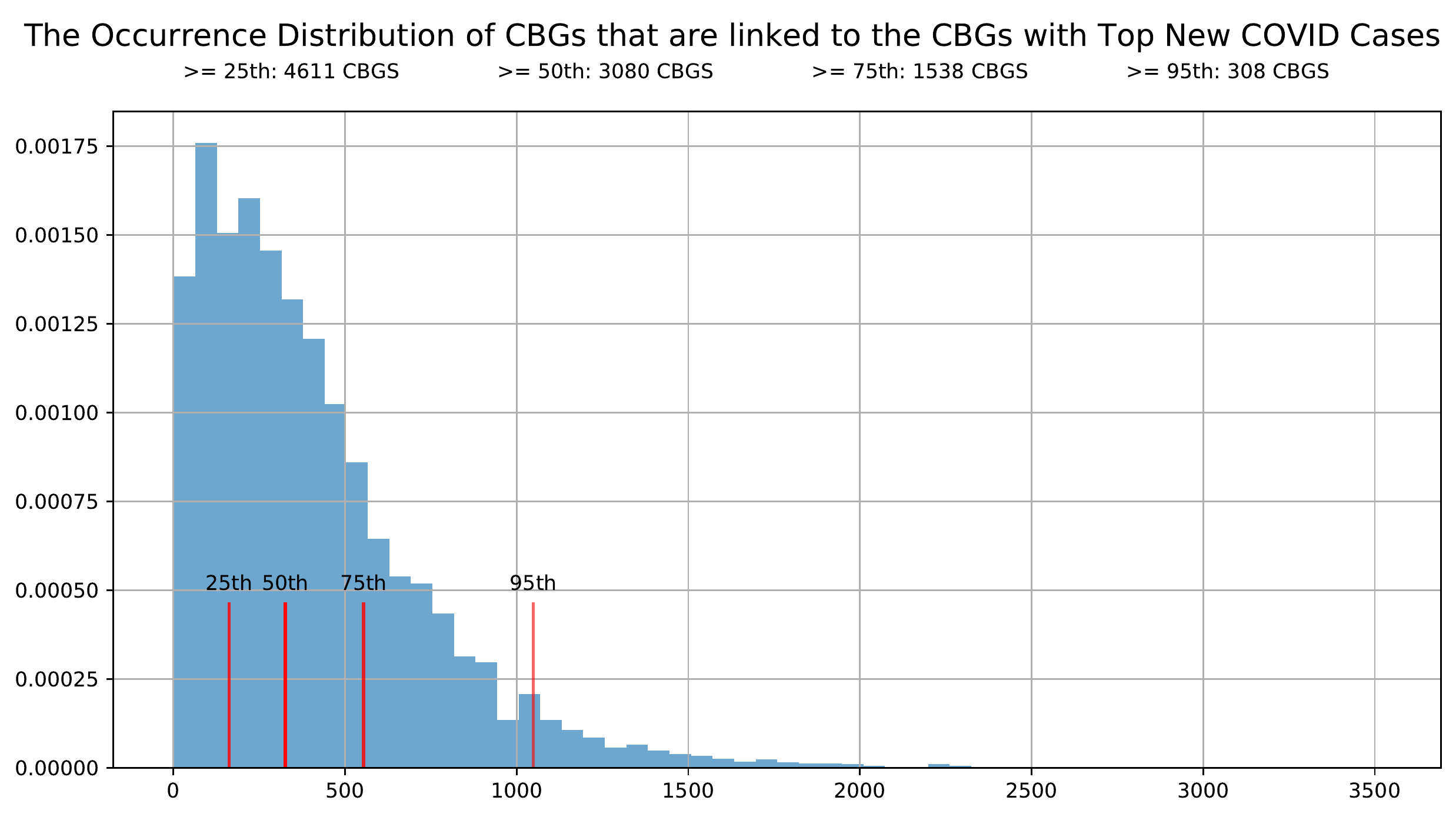}
\caption{Bridge occurrence distribution with percentiles. CBGs are ranked with respect to how frequently they appear in the neighborhood of COVID hotspots in a weekly manner. We use occurrence percentiles and consider the ones above the 75$^{th}$ percentile as the final bridge CBGs group.}
\label{fig:hotspot-dist}
\end{figure}

\newpage

\begin{figure}[hbt!]
\centering
\includegraphics[width=\linewidth]{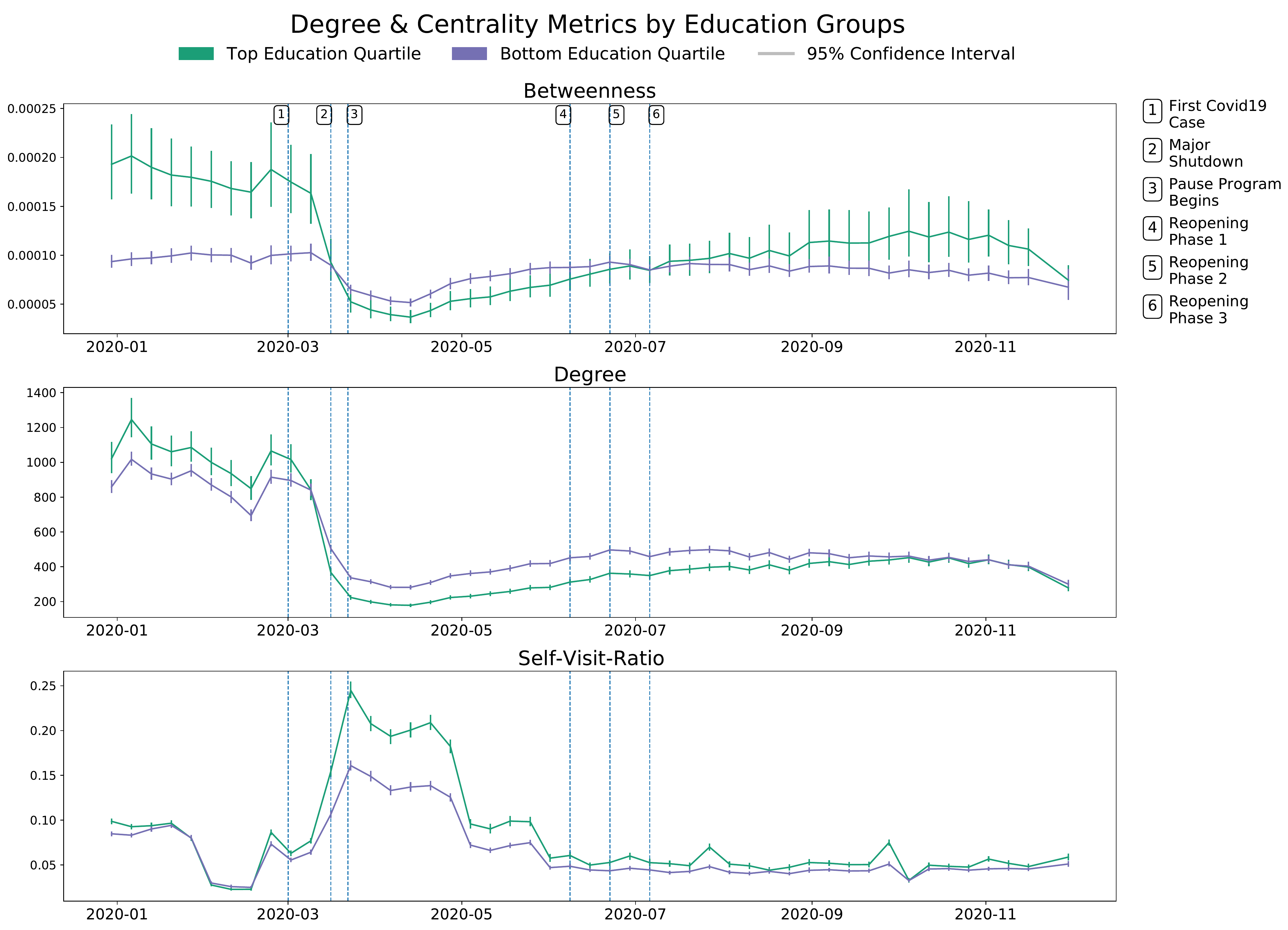}
\caption{Centrality metrics in top and bottom education quartiles. The figure displays similar trends between top and bottom quartiles as in the income groups in Figure 2 of the main paper. With the emergence of the COVID and the following curfews and lockdowns, roles of the bottom and top quartiles are flipped for betweenness and degree metrics. For self-visit-ratio, the difference between the groups drifted even more.}
\label{fig:education-cent}
\end{figure}

\newpage

\begin{figure}[hbt!]
\centering
\includegraphics[width=\linewidth]{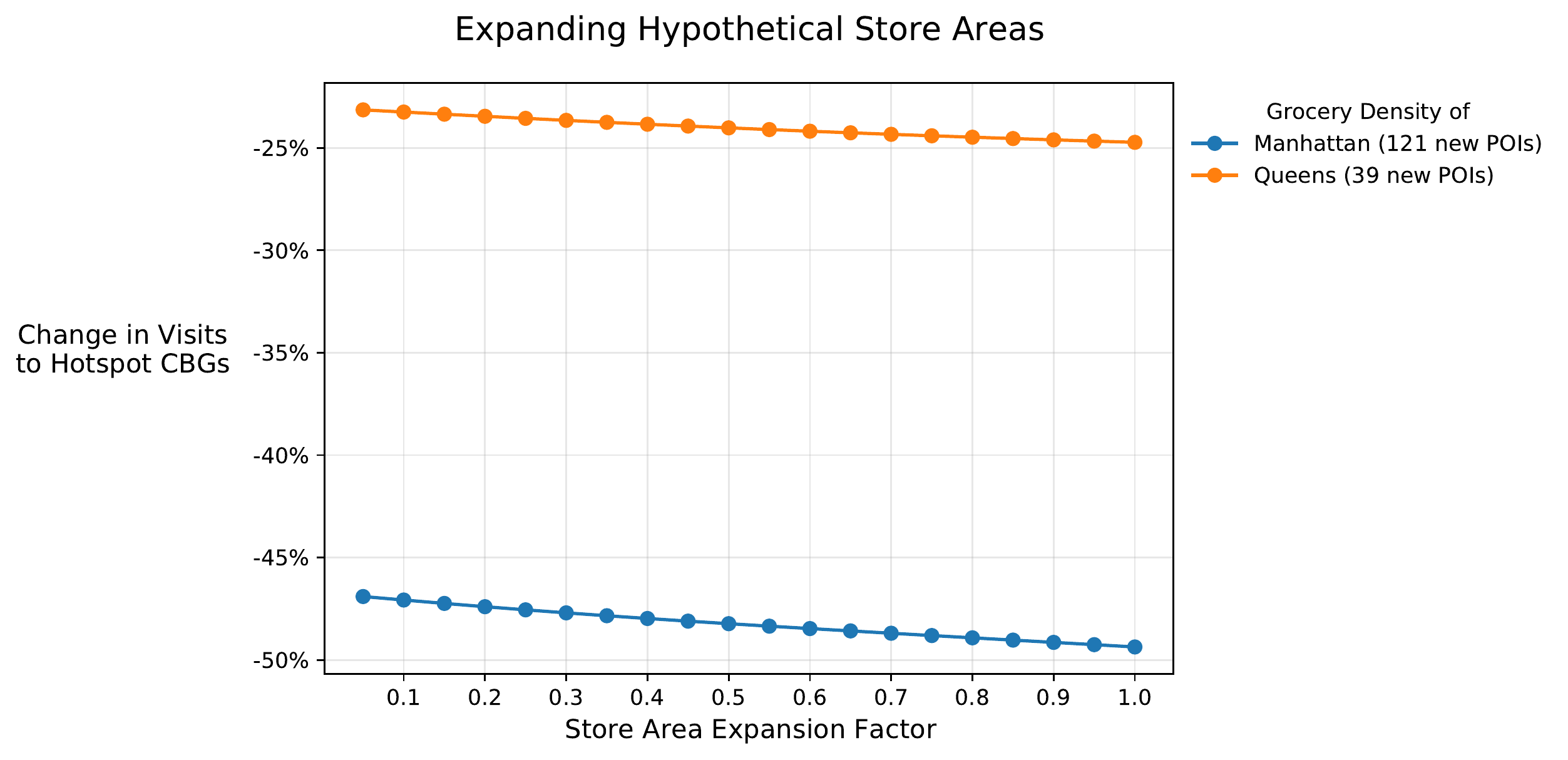}
\caption{Change in visits to hotspot CBGs in Staten Island with the hypothetical POI density matching Manhattan and Queens. Initially, the mean POI size in Staten Island is used for the hypothetically generated POIs. Each dot represents a setting with target borough's POI density in which hypothetical POI size is incremented by the expansion factor.}
\label{fig:store-area-huff}
\end{figure}

\newpage

\begin{figure}[hbt!]
\centering
\includegraphics[width=\linewidth]{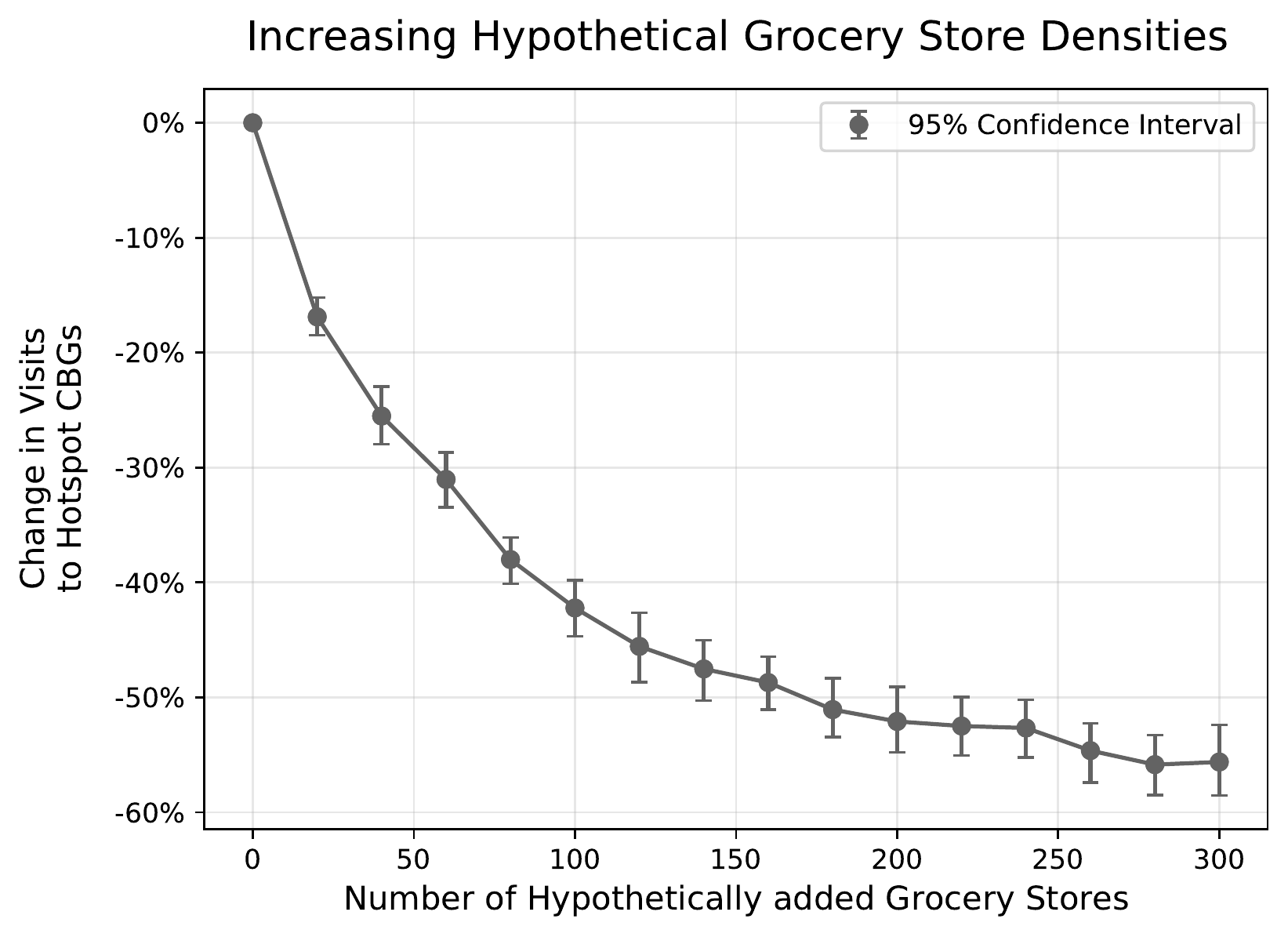}
\caption{Change in visits to hotspot CBGs in Staten Island with different number of hypothetical POI additions. In contrast to the POI area expansion, the addition of POIs display rapid decrease in visits to hotpot CBGs.}
\label{fig:num-store-huff}
\end{figure}

\newpage

\begin{figure}[hbt!]
\centering
\includegraphics[width=\linewidth]{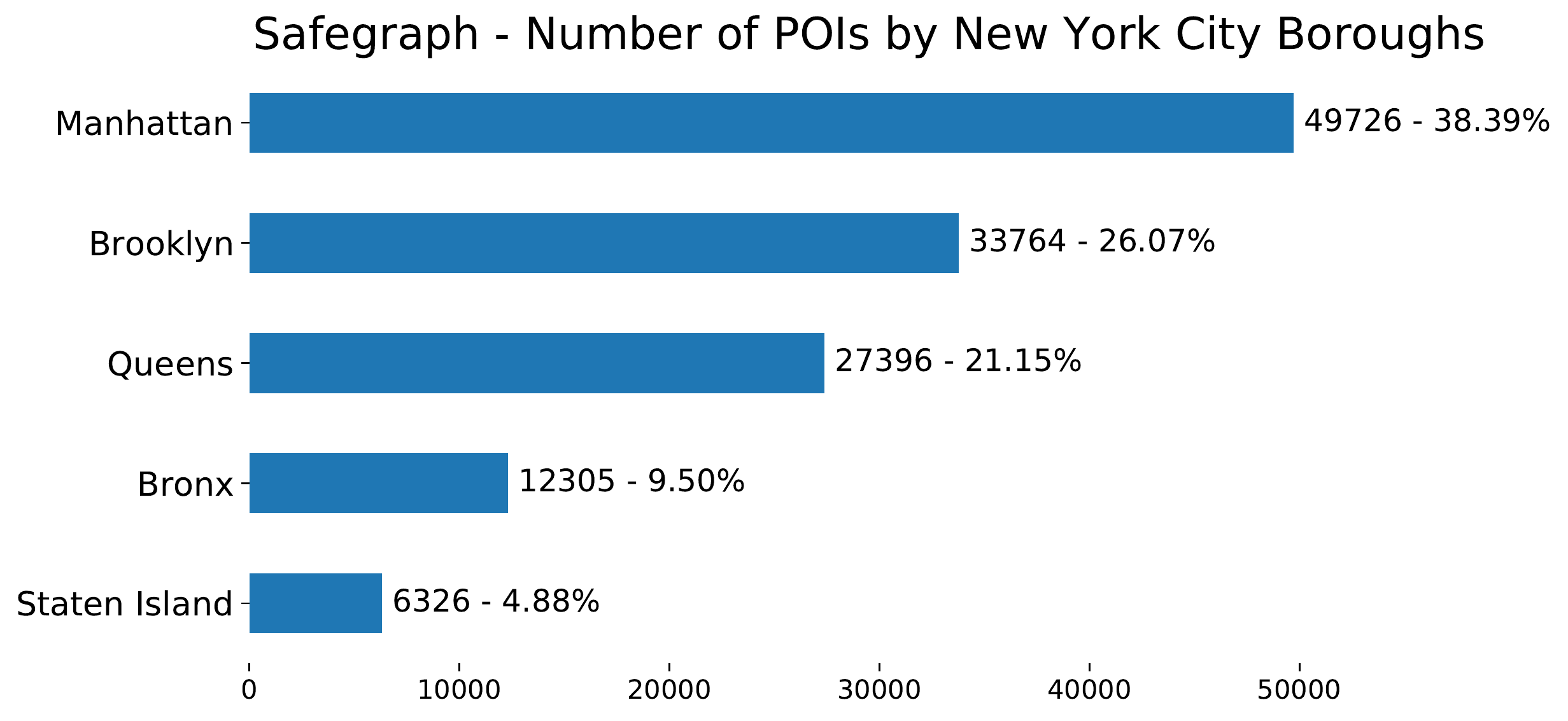}
\caption{Number of POIs in each borough in the Safegraph dataset. Staten Island has the lowest share of POIs in New York City.}
\label{fig:poi-dist-borough}
\end{figure}

\bibliographystyle{ieeetr}
\bibliography{supplement}